\documentclass[aps,pre,twocolumn,showpacs]{revtex4}
\usepackage{graphics,graphicx,subfigure}
\usepackage{amsmath}

\newtheorem{thm}{Theorem}

\newtheorem{example}{Example}

\newcommand{\rank}{\mathrm{rank\,}}
\newcommand{\Tr}{\mathrm{\,Tr\,}}

\newcommand{\ben}{\begin{enumerate}}
\newcommand{\een}{\end{enumerate}}
\newcommand{\bit}{\begin{itemize}}
\newcommand{\eit}{\end{itemize}}
\newcommand{\be}{\begin{equation}}
\newcommand{\ee}{\end{equation}}
\newcommand{\bdm}{\begin{displaymath}}
\newcommand{\edm}{\end{displaymath}}
\newcommand{\bea}{\begin{eqnarray}}
\newcommand{\eea}{\end{eqnarray}}
\newcommand{\f}[1]{\fbox}

\newcommand{\realnos}{\mbox{{\bf R}}}

\newcommand{\integers}{\mbox{{\bf Z}}}

\begin{document}

\title{Temporal Behavior of the Conditional and Gibbs' Entropies}

\author{Michael C. Mackey}
\email{mackey@cnd.mcgill.ca} \affiliation{Departments of Physiology, Physics \& Mathematics and Centre for
Nonlinear Dynamics, McGill University, 3655 Promenade Sir William Osler, Montreal, QC, CANADA, H3G 1Y6}
\author{Marta Tyran-Kami\'nska}
\email{mtyran@us.edu.pl}
\thanks{Corresponding author}
\affiliation{Institute of Mathematics, Silesian University, ul. Bankowa 14, 40-007 Katowice, POLAND}
\date{\today}

\begin{abstract}
We study the  temporal  approach  to equilibrium of the
Gibbs' and  conditional entropies for both invertible
deterministic dynamics as well as non-invertible stochastic
systems in the presence of white noise.  The conditional
entropy will either remain constant or monotonically
increase to its maximum of zero. However, the Gibbs'
entropy may have a variety of patterns of approach to its
final value ranging from a monotone increase or decrease to
an oscillatory approach.  We have illustrated all of these
behaviors using examples in which both entropy dynamics can
be determined analytically.

  \end{abstract}
    \maketitle

\section{Introduction}\label{s:intro}

A variety of various measures of dynamic behavior carry the
name of entropy. Two have proved to be especially
intriguing in the examination of the temporal evolution of
dynamical systems when considered from an ensemble point of
view.

One of these is known as the conditional entropy.
Convergence properties of the conditional entropy have been
extensively studied because `entropy methods'  have been
known for some time to be useful for problems involving
questions related to convergence of solutions in partial
differential equations
\citep{loskot91,abbond99,toscanivillani,arnold01,qian01,qian02}.
Their utility can be traced, in some instances, to the fact
that the conditional  entropy may serve as a Liapunov
functional \citep{markowich00}.

Another type of entropy is the Gibbs' entropy, which is
strongly related to an extension of the equilibrium entropy
that was introduced by \citet{gibbs02} to a time dependent
situation.  This has been considered by a number of authors
recently \citet{ruelle96,ruelle03}, \citet{nicolis98},
\citet{nicolis99} and
\citet{bag00,bag01,bag02,bag02b,bag03}.

Here we compare and contrast the temporal evolution of the
conditional and Gibbs' entropies in a variety of dynamical
settings.  Our primary considerations are stochastic
non-invertible systems with  additive white noise, but we
also do discuss the two entropy behaviors in systems with
invertible dynamics.

The organization of the paper is as follows.
Section~\ref{s:prelim} gives some basic background,
definition of steady state Gibbs' entropy, and extension of
this to time dependent situations, and shows how the
conditional entropy can be considered a generalization of
the time dependent Gibbs' entropy. Section~\ref{s:det}
looks at the behavior of the Gibbs' entropy and the
conditional entropy in systems with invertible dynamics.
The general results of this section are illustrated with
three specific examples. This is extended to non-invertible
systems in Section~\ref{s:noninvert} where we cite a number
of results from \citep{mackeytyran05} on the behavior of
the conditional entropy and contrast these with the
behavior of the Gibbs' entropy. These considerations are
illustrated with two detailed examples drawn from dynamical
systems perturbed by noise. The paper concludes with a
summary and discussion in Section~\ref{s:disc} where we
show how  our results illuminate the connection between a
previously postulated dynamic analog of the non-equilibrium
thermodynamic entropy and the  entropy increase of  the
second law of thermodynamics.

\setcounter{equation}{0}
\section{Gibbs' and conditional Entropies}\label{s:prelim}

Let $X$ be a phase space and $\mu$  a reference measure on
$X$.  Denote the corresponding
set of densities by $\mathcal{D}(X)$, or $\mathcal{D}$ when
there will be no ambiguity, so $f \in \mathcal{D}$ means $f
\geq 0$ and $\int_X f(x) \,dx = 1$ (for integrals with respect to the reference measure we
use the notation $\int f(x)\,dx$ rather than $\int
f(x)\,d\mu(x)$). Let $\{P^t\}_{t\ge 0}$ be a semigroup of
Markov operators on $L^1(X)$, {\it i.e.} $P^tf_0 \ge 0$ for an initial density
$f_0\ge 0$,
 $\int P^tf_0 (x)\, dx=\int f_0 (x) \,
dx$, and $P^{t+s}f_0=P^t(P^sf_0)$. If the group property
holds for $t,s \in \realnos$, then we say that $P^t$ is
invertible. If  it holds only for $t,s \in \realnos ^+$ we
say that $P^t$ is non-invertible.  If there is a density
$f_*$ such that $P^tf_*=f_*$ for all $t>0$, $f_*$ is called
a stationary density of $P^t$.


In his seminal work Gibbs \cite{gibbs02},  assuming the
existence of a system steady state  density $f_*$ on the
phase space ${ X}$, introduced the concept of the index of
probability given by $\log f_*(x) $ where ``$\log$" denotes
the natural logarithm.  He then identified the entropy in a
steady state  situation with the average of the index of
probability
\begin{equation}
H_G(f_*) = - \int_{ X}  f_*(x) \log f_*(x)\, dx,
\label{e-gibbs}
\end{equation}
and we call this the {\it equilibrium or steady state
Gibbs' entropy}.

If entropy is to be an extensive quantity   (in accord with
experimental evidence) then Definition~\ref{e-gibbs} is
unique up to a multiplicative constant
\cite{khinchin49,skagerstam74}. It is for this reason that
we extend the definition of the steady state  Gibbs'
entropy to time dependent (non-equilibrium) situations and
say that the {\it time dependent Gibbs' entropy} of a
density $f(t,x)$ is defined by
\begin{equation}
H_G(f) = - \int_{ X}  f(t,x) \log f(t,x)\, dx.
\label{e-gibbstime}
\end{equation}

We define the conditional entropy as   \citep{almcmbk94}
    \be
H_c(f|f_*)=-\int_X f(t,x)\log\dfrac{f(t,x)}{f_*(x)}dx.
\label{d:conent}
    \ee
It is variously known as 
the Kullback-Leibler or relative entropy \citep{loskot91},
the relative Boltzmann entropy \citep{eu95,eu97}, or the
specific relative entropy \citep{jiang00}, and has been
related to the free energy \citep{qian01,qian02,qian02a}.
If there is a convergence $\lim_{t \to \infty} f(t,x) =
f_*(x)$ in some sense (which we will make totally precise
in Section~\ref{s:noninvert}) then $\lim_{t \to \infty}
H_c(f|f_*)= 0$.

\setcounter{equation}{0}
\section{Entropy behavior and invertible dynamics}
\label{s:det}

This section considers the behavior of the   Gibbs' entropy
and the conditional entropy in situations where the
dynamics are invertible in the sense that they can be run
forward or backward in time without ambiguity. To make this
clearer, consider a phase space $X$ and a dynamics $S_t:X
\to X$. For every initial point $x_0$, the sequence of
successive points $S_t(x_0)$, considered as a function of
time $t$, is called a trajectory. In the phase space $X$,
if the trajectory $S_t(x_0)$ is nonintersecting,  or
intersecting but periodic, then at any given final time
$t_f$ such that $x_f = S_{t_f}(x_0)$ we could change the
sign of time by replacing $t$ by $-t$, and run the
trajectory backward using $x_f$ as a new initial point in
$X$. Then our new trajectory $S_{-t}(x_f)$ would arrive
back at $x_0$ after a time $t_f$ had elapsed: $x_0 =
S_{-t_f}(x_f)$.  Thus in this case we have a dynamics that
may be reversed in time {\it completely unambiguously}.
Dynamics with this character are known variously as time
reversal invariant \citep{sachs87}, or reversible
\citep{reichenbach57}  in the physics literature, and as
invertible in the mathematics literature \citep{almcmbk94}.

We formalize this by introducing the concept of a dynamical
system $\lbrace S_t \rbrace _{t \in \realnos}$, which is
simply any group of transformations $S_t:X \rightarrow X$
having the two properties: 1. $S_0 (x) = x$; and 2.
$S_t(S_{t'}(x)) = S_{t+t'}(x)$ for $t,t'\in \realnos $ or
$\integers$. Since, from the definition, for any $t \in
\realnos$, we have $S_t(S_{-t}(x)) = x = S_{-t}(S_t(x))$,
dynamical systems are invertible in the sense discussed
above since they may be run either forward or backward in
time. Systems of ordinary differential equations are
examples of dynamical systems.

Our first result from \cite{mackeytyran05}  shows that the conditional entropy of any
invertible system is  uniquely determined by the  system
preparation and does not change with time.  This is
formalized in

\begin{thm}(\citep[Theorem 3]{mackeytyran05})
\label{thm-invert} If $P^t $ is an invertible Markov
operator and has a stationary density $f_*$, then the
conditional entropy is constant  and equal to the value
determined by $f_*$ and the choice of the initial density
$f_0$ for all time $t$. That is,
\begin{equation}\nonumber
H_c (P^tf_0  |f_*) \equiv H_c(f_0|f_*) 
\end{equation}
for all $t$.
\end{thm}

 More specifically, when  considering a deterministic dynamics $S^t: { X}
\to { X}$, the corresponding Markov operator is also known
as the Frobenius Perron operator \citep{almcmbk94} and is
given by
    \be
    P^tf_0 (x) = f_0(S^{-t}(x)) | J^{-t}(x)|,
    \ee
where $J^{-t}(x)$ denotes the Jacobian of $S^{-t}(x)$.
 Further,
    \begin{equation}\nonumber
    \begin{split}
    H_c(P^tf_0 |f_*) &= -\int_ { X} P^t f_0(x)\log \left [ \dfrac{P^t f_0(x)}{f_*(x)}\right ] dx  \\
    &= -\int_ { X} \dfrac{f_0(S^{-t}(x))}{|J^{t}(x)|}\log \left [\dfrac{ f_0(S^{-t}(x)) }{f_*(S^{-t}(x))}\right ] dx  \\
    &= -\int_ { X} f_0(y)\log\left [ \dfrac{ f_0( y) } {f_*(y)}\right ] dy  \\
    &\equiv  H_c(f_0|f_*)
       \end{split}
    \end{equation}
as expected from Theorem~\ref{thm-invert}. This behavior
is, however, quite different from what is seen in the
  Gibbs' entropy since
    \begin{equation}\nonumber
    \begin{split}
    H_{G}(P^tf_0 ) &= -\int _{ X} P^tf_0(x) \log [P^tf_0(x)]\,dx  \\
    &= -\int _{ X} \dfrac{f_0(S^{-t}(x))}{ |J^{t}(x)|}  \log \left[\dfrac{f_0(S^{-t}(x))}{ |J^{t}(x)|}\right]dx  \\
    &= -\int _{ X} f_0( y) \log \left [ \dfrac {f_0(y)}{ |J^{t}(y)| } \right ]dy  \\
    &= H_{G}(f_0) + \int _{ X} f_0( y) \log  |J^{t}(y)|dy.
    \end{split}
    \end{equation}
Thus, in spite of the fact that the conditional entropy is
constant for an invertible system, the   Gibbs' entropy may
continually change, and satisfies
    \be
    H_{G}(P^tf_0 ) - H_{G}(f_0) = \int _{ X} f_0( y) \log
    |J^{t}(y)|dy.
    \ee
Note in particular that for the   Gibbs' entropy to be an
increasing function of time,
we must have an expanding
dynamics in the sense that
    \be
    \int _{ X} f_0( y) \left \{ \log
    |J^{t}(y)| - \log|J^{t'}| \right \} dy > 0 \qquad
    \mbox {for} \qquad t > t'.
    \ee

When the Jacobian is constant,
    \be
    H_{G}(P^tf_0 ) - H_{G}(f_0) = \log  |J^{t}|,
    \ee
illustrating that the   Gibbs' entropy $H_{G}(P^tf_0 )$ may either deviate
from or approach the initial entropy    $H_{G}(f_0)$  depending on the value
of $|J^{t}|$. If the Lebesgue measure is preserved so
$|J^t| = 1$, then $H_{G}$ is constant.

Taking an even more specific example, if the dynamics corresponding to the invertible Markov
operator are described by the system of ordinary
differential equations
    \be
    \dfrac{dx_i}{dt} = F_i(x) \qquad        i = 1,\ldots ,d
    \label{ode}
    \ee
operating in a region  ${ X} \subset \realnos^d$ with
initial conditions $x_i(0) = x_{i,0}$, then
\citep{almcmbk94} the evolution of $f(t,x) \equiv P^tf_0
(x)$ is  governed by the generalized Liouville  equation
\begin{equation}
\frac {\partial f}{\partial t} = -\sum_i \frac {\partial (fF_i)}{\partial x_i}. \label{e-leqn}
\end{equation}
If the stationary density  $f_*$ exists, it is given by the
solution of
\begin{equation}
\sum_i \frac {\partial (f_* F_i)}{\partial x_i} = 0. \label{e-liouss}
\end{equation}
Note that the constant function 
$f_* \equiv 1$, meaning that the flow defined by
Eq.~\ref{ode} preserves the Lebesque measure, is a
stationary solution of Eq.~\ref{e-leqn} if and only if
\begin{equation}
\sum_i \frac {\partial F_i}{\partial x_i} = 0, \label{e-liouss1}
\end{equation}
but if $X$ has an infinite Lebesgue measure then $f_*$ is
not integrable, so there will be no stationary density.

The rate of change of the   Gibbs' entropy is given by \be
\dfrac {dH_{G}}{dt} = \int_{ X} f \sum_i \frac {\partial
F_i}{ \partial x_i} \,dx,\label{gibbsrate} \ee and so it is
only in Lebesgue measure preserving dynamics (like
Hamiltonian systems), for which Eq.~\ref{e-liouss} holds,
that $H_{G}$ will be constant.  This was first noted by
\citet[pp. 143-4]{gibbs02} and much later pointed out in
\citep{steeb79} and proved in general in \citep{andrey85},
as emphasized in \citep{ruelle03}. If
    \be
    \sum_i \frac {\partial F_i }{ \partial x_i} > 0,
    \ee
    then the   Gibbs' entropy will increase.

\begin{example}\label{exp:ct-friction}To illustrate these
points, consider the  continuous time dynamical system on
$\realnos^2$
    \be
    \dfrac {dx}{dt} = F x 
\label{eq:ct}
    \ee
where $F=(F_{ij})$ is a $2\times 2$ matrix.  It can be
solved exactly and its solution is given by $$
x(t)=e^{tF}x(0),\qquad \mbox{where}\qquad
e^{tF}=\sum_{n=0}^\infty \frac{t^n}{n!}F^n.
$$
The evolution of the density $f$ under the action of this
flow is determined by the solution of the Liouville
equation
    \be
    \dfrac{\partial f}{\partial t} = -\dfrac{\partial (F_1(x)f)}{\partial
    x_1} -\dfrac{\partial (F_2(x)f)}{\partial
    x_2},\label{e:friction-fp}
    \ee
where $F_j(x)=F_{j1}x_1+F_{j2}x_2$, $j=1,2$.  If the
initial density is given by $f_0(x)$, then the general
solution of Eq.~\ref{e:friction-fp} is given by
$$
f(t,x) = |\det e^{-tF}|f_0(e^{-tF}x).
$$
Since
$$
|\det e^{-tF}|=e^{-t \Tr F },
$$
where $\Tr F=F_{11}+F_{22}$ is the trace of the matrix $F$,
we have
$$
P^tf_0 (x) = e^{-t\Tr F}f_0(e^{-tF}x).
$$
Consequently, we obtain
$$
H_G(P^tf_0 ) = H_G(f_0) + t \Tr F .
$$
If $\lambda_1,\lambda_2$ are the eigenvalues of $F$ then
$\Tr F=\lambda_1+\lambda_2$. Thus when $\Tr F < 0$ then the
system has a one dimensional attractor and  $ H_G(P^tf_0
)\to -\infty$ as $t\to\infty$, while if $\Tr F
> 0$ then the dynamics are sweeping \cite{almcmbk94} .
However, in this example there will be no stationary
density and the conditional entropy is thus not defined.
\end{example}

\begin{example}\label{exp:ct-ho}
Consider the second order system
    \be
    m\dfrac{d^2y}{dt^2}+ \gamma \dfrac{dy}{dt}+\omega^2 y=0
    \label{e:b-osca}
    \ee
with constant coefficients $m$, $\gamma$ and $\omega$.
Introduce the velocity $v= \dot y$  as a new variable. Then
Eq.~\ref{e:b-osca} is equivalent to the system
    \begin{subequations}\label{e:b-osc}
    \begin{align}\label{e:b-osc1a}
    \dfrac{dy}{dt} &= v \\
    m \dfrac{dv}{dt} &= -\gamma v-\omega^2 y ,\label{e:b-osc2a}
    \end{align}\end{subequations}
thus by writing
\begin{equation}\nonumber
 x=\left(
      \begin{array}{c}
       y   \\
       v
      \end{array}
    \right)
\qquad \mbox{and}\qquad F=\left(\begin{array}{cc} 0 & 1\\
-\dfrac{\omega^2}{m}& -\dfrac{\gamma}{m}
\end{array}\right)
 \label{e:b-osc2am}
    \end{equation}
we recover Eq.~\ref{eq:ct}. Since $\Tr F=-\gamma/m$, we
obtain
$$
P^tf_0 (x)=e^{\gamma t/m}f_0(e^{-tF}x)
$$
and
$$
H_G(P^tf_0 )=H_G(f_0)-\dfrac{\gamma t}{m}.
$$
As in Example~\ref{exp:ct-friction}, when there is damping
so $\Tr F= -\gamma/m<0$, then $ H_G(P^tf_0 ) \to
-\infty$ as $t\to\infty$. Further, as in the previous
example there is no stationary density $f_*$ and so
$H_c(f|f_*)$ is not defined.
\end{example}

\begin{example}\label{exp:rotation}
Let $X$ be the unit circle in $\realnos^2$. If
$$
F=\left(
    \begin{array}{rr}
      0 & 1 \\
    -1 & 0 \\
\end{array}
\right)
$$
then Eq.~\ref{eq:ct} has the general solution
$$
x(t)=\left(
    \begin{array}{rr}
      \cos t & \sin t \\
    -\sin t & \cos t \\
\end{array}
\right)x(0)
$$
and $x(t)\in X$ for $x(0)\in X$. If $\mu$ is the Lebesgue measure on
$X$ then the corresponding Perron-Frobenius operator is given by
$$P^tf_0 (x)=f_0(x(-t))\quad \mbox{ for }\quad f_0\in L^1(X)$$ and
$f_*(x)=\dfrac{1}{2\pi}1_X(x)$ is the stationary density of $P^t$.
We then have
\begin{equation}\nonumber
\begin{split}
H_c(P^tf_0 |f_*)&=H_G(P^tf_0)-\log 2\pi\\
&=H_G(f_0)-\log 2\pi=H_c(f_0|f_*)
\end{split}
\end{equation}
and both entropies are constant and fixed by the initial
system preparation ($f_0$).
\end{example}

\setcounter{equation}{0}
\section{Entropy behavior and non-invertible dynamics}\label{s:noninvert}

\subsection{Asymptotic stability and conditional entropy}\label{ss:asre}

A semigroup of Markov operators $P^t$ on $L^1(X)$ is said
to be {\it asymptotically stable} if there is a stationary
density $f_*$ of $P^t$ such that for all initial densities $f_0$
$$
\lim_{t\to\infty}P^tf_0 =f_*
$$
(here the limit denotes convergence in $L^1(X)$). Systems
with dynamics that are asymptotically stable must, by
necessity, be non-invertible \citep[Remark
4.3.1]{almcmbk94}.

\begin{thm}(\citep{voigt81})\label{t:voigt} Let  $P^t$ be a semigroup of Markov operators on $L^1(X)$ and
$f_*$ be a stationary density. Then for every density $f_0$ the
conditional entropy $H_c(P^tf_0 |f_*)$ is a nondecreasing function of
$t$.
\end{thm}

For a given density $f_0$  the conditional entropy $H_c(P^tf_0 |f_*) $ is
bounded above by zero. Thus we know that it has a limit as
$t\to\infty$. Our next result connects the temporal convergence
properties of $H_c$ with those of $P^t$.

\begin{thm}(\citep[Theorem 1]{mackeytyran05})
\label{t:entropyconv}
Let  $P^t$ be a semigroup of Markov operators on $L^1(X)$
and $f_*$ be a stationary density. Then
    $$
    \lim_{t \to \infty} H_c(P^tf_0 |f_*) = 0
        $$
for all $f_0$ with $H_c(f_0|f_*)>-\infty$ if and only if $P^t$ is
asymptotically stable.
\end{thm}

A consequence of the  convergence of the  conditional
entropy to zero is that
 \be\nonumber
\lim_{t\to\infty}\int h(x)P^tf_0 (x)\,dx=\int
h(x)f_*(x)\,dx
 \ee for any measurable function $h$ for which the integral
 \be \nonumber\int
 e^{rh(x)}f_*(x)\,dx
 \ee is finite for all $r$ in some
neighborhood of zero \citep[Lemma 3.1]{csiszar75}. Since
the conditional and Gibbs' entropies are related by
$$
H_G(P^tf_0 )=H_c(P^tf_0 |f_*)-\int P^tf_0(x)\log f_*(x)
\,dx,
$$
 Theorem~\ref{t:entropyconv} implies
\begin{thm}\label{th-convg}
Let  $P^t$ be an asymptotically stable semigroup of Markov
operators on $L^1(X)$ with a stationary density $f_*$ such
that $\displaystyle\int f_*^{1+r}(x)\, dx <\infty$ for all
$r$ in some neighborhood of zero. Then
$$
\lim_{t\to\infty}H_G(P^t f_0)=H_G(f_*)
$$
for all $f_0$ with $H_c(f_0|f_*)>-\infty$.
\end{thm}

Theorem~\ref{t:entropyconv} shows that 
asymptotic stability is necessary and sufficient for the
convergence of $H_c$ to zero. A weaker property for entropy
convergence has been previously considered by
\citet{ruelle76,ruelle80} and later popularized
\citep{evans90,evans93} as a `chaotic hypothesis'
\citep{gallavotti}.  The chaotic hypothesis postulates that
underlying dynamics are (invertible) Anasov systems.

\subsection{Effects of noise in continuous time systems}\label{s:gauss}

In  this section, we consider  the behavior of the entropies $H_G(P^tf_0 )$ and  $H_c(P^tf_0|f_*)$ when the dynamics are described by the  stochastically
perturbed system
    \be
    \dfrac{dx_i}{dt} = F_i(x) + \sum_{j=1}^d \sigma_{ij}(x) \xi _j, \qquad   i = 1,\ldots ,d
    \label{stochode}
    \ee
with  the initial conditions $x_i(0) = x_{i,0}$.   $\sigma
_{ij}(x)$ is the amplitude  of the stochastic perturbation
and  $\xi_j = \dfrac {dw_j}{dt}$ is a white noise  term
that is the derivative of a Wiener process. It is  assumed
that the It\^o, rather that the Stratonovich, calculus, is
used.  (For  the differences see \cite{horsthemke84},
\cite{almcmbk94} and \cite{risken84}. If  the $\sigma_{ij}$
are independent of $x$ then the It\^o and  the Stratonovich
approaches yield identical results.)

The {\it  Fokker-Planck equation}   governing the evolution
of the density function $f(t,x)$  is given by
    \be
    \frac {\partial f}{\partial t} = -  \sum_{i=1}^d \frac{\partial
    [F_i(x)f]}{\partial x_i} +\frac 12 \sum_{i,j=1}^d \frac{\partial ^2
    [a_{ij}(x)f]}{\partial x_i \partial x_j}
    \label{fpeqn}
    \ee
where
$$
a_{ij}(x)=\sum_{k=1}^{d}\sigma_{ik}(x)\sigma_{jk}(x).
$$
If $k(t,x,x_0)$  is the fundamental solution of the
Fokker-Planck equation, i.e. for every $x_0$ the function
$(t,x)\mapsto k(t,x,x_0)$ is a solution of the
Fokker-Planck equation with the initial condition
$\delta(x-x_0)$, then the general solution $f(t,x)$ of the
Fokker-Planck equation (\ref{fpeqn}) with the initial
condition $ f(x,0)=f_0(x)$ is given by \be f(t,x)=\int
k(t,x,x_0)f_0(x_0)\, dx_0,
\label{gensoln}
\ee and defines a Markov semigroup by $ P^tf_0(x)=f(t,x). $

If a stationary (steady state)  density  $f_*(x)$ exists,
it is the stationary solution of  Eq.~\ref{fpeqn}:
    \be
    -  \sum_{i=1}^d \frac{\partial
    [F_i(x)f]}{\partial x_i} +\frac 12 \sum_{i,j=1}^d \frac{\partial ^2
    [a_{ij}(x)f]}{\partial x_i \partial x_j} = 0.
    \label{ssfpeqn}
        \ee

Differentiating Eq.~\ref{e-gibbstime} with respect to time,
and using Eq.~\ref{fpeqn} with integration by parts along
with the the fact that since $f_*$ is a stationary density
it satisfies (\ref{ssfpeqn}), we obtain
    \be
    \dfrac{dH_c}{dt} = \dfrac12 \int \left( \dfrac{f_*^2}{f} \right)
    \sum_{i,j=1}^{d} a_{ij}(x) \dfrac {\partial}{\partial x_i} \left( \dfrac {f}{f_*} \right)
    \dfrac {\partial}{\partial x_j} \left( \dfrac{f}{f_*} \right) \,dx.
    \label{dtcondent}
    \ee
    Since the matrix $(a_{ij}(x))$ is nonnegative definite, one
concludes that $\dfrac{dH_c}{dt}\ge 0$. Using the identity
$$
\dfrac
{\partial}{\partial x_i}\left( \log \dfrac {f}{f_*}
\right)=\dfrac {f_*}{f}\dfrac {\partial}{\partial x_i}
\left(\dfrac {f}{f_*}\right),
$$
we can rewrite Eq.~\ref{dtcondent} in the equivalent form 
    \be
    \dfrac{dH_c}{dt}=
   \dfrac12 \int f
    \sum_{i,j=1}^{d} a_{ij}(x) \dfrac {\partial}{\partial x_i} \left(\log \dfrac {f}{f_*} \right)
    \dfrac {\partial}{\partial x_j}\left(\log \dfrac{f}{f_*} \right) \,dx.
\label{e:dtcondent}
    \ee
The right hand side of Eq.~\ref{e:dtcondent} appears in
various expressions describing entropy balance equations in
\citet{nicolis99} and \citet{bag02}.

A similar calculation for the   Gibbs' entropy yields, however,
something additional. Namely, we have
\begin{widetext}
    \be
    \dfrac{dH_{G}}{dt} = \int  f \left(\sum_i \frac
    {\partial F_i(x)}{\partial x_i} - \dfrac12  \sum_{i,j} \frac
    {\partial^2 a_{ij}(x)}{\partial x_i \partial x_j}\right) \,dx +
    \dfrac12 \int \dfrac{1}{f}
    \sum_{i,j=1}^d a_{ij}(x)\dfrac
    {\partial f}{\partial x_i} \dfrac
    {\partial f}{\partial x_j}\, dx.
       \label{bgnoise}
    \ee
    \end{widetext}
If $a_{ij}$ are independent of $x$ then we obtain
    \be
    \dfrac{dH_{G}}{dt} = \int  f \sum_i \frac
    {\partial F_i(x)}{\partial x_i}  \,dx +
    \dfrac12 \sum_{i,j=1}^d a_{ij}\int \dfrac{1}{f}
    \dfrac
    {\partial f}{\partial x_i} \dfrac
    {\partial f}{\partial x_j}\, dx.
       \label{c-bgnoise}
    \ee

As pointed out in \citep{nicolis99}, the first term is of
indeterminant sign, while the second is positive definite
so the temporal behavior of the Gibbs' entropy in this
non-invertible system is unclear.  It has become customary
\citep{nicolis99,bag00,bag01,bag02,bag02b,bag03,bag04} to
refer to the first term in Eq.~\ref{bgnoise} as the
`entropy flux' and the second term as the `entropy
production'.

There are a number of results giving conditions such that
the  general solution $f(t,x)$ of the Fokker Planck
equation is asymptotically stable and thus the conditional
entropy evolves monotonically to zero, e.g. 
Theorem 11.9.1 in \cite{almcmbk94}. In
particular, assume that the stationary density is of the
form
$$
f_*(x)=e^{-B(x)}.
$$
From Theorem~\ref{t:entropyconv} it follows that $$\lim_{t
\to \infty}H_c(f |f_*) = 0$$ and from
Theorem~\ref{th-convg} that
$$\lim_{t \to \infty}H_G(f ) = H_G(f_*)$$
for all $f_0$ with $H_c(f_0|f_*)>-\infty$ provided that
$\int e^{-(1+r)B(x)}dx<\infty$ for $r$ in some neighborhood
of zero.

From the definition of the conditional entropy
we may write
     \be
    H_c(f|f_*) = H_G(f) + \int_X f(t,x)\log f_*(x) dx \label{d:con-bg-ent1} 
    \ee
so the derivative of the Gibbs' entropy is
 \be
 \dfrac{dH_{G}}{dt} =  \dfrac{dH_{c}}{dt} - \int Lf \log f_*
dx, \ee where the operator $L$ is given by \be
    Lf=  -\sum_{i=1}^d  \frac{\partial(F_i(x)f) }{\partial x_i} +\frac
    {1}{2} \sum_{i,j=1}^d  \frac{\partial ^2 (a_{ij}(x)f)}{\partial x_i \partial
x_j}.
    \label{l}
    \ee
Since $\log f_*(x)=-B(x)$, we may write
 \be
\dfrac{dH_{G}}{dt}=\dfrac{dH_{c}}{dt} + \int f L^-B
dx,\label{con-bgnoise-d}  \ee where the operator $L^-$ is
the formal adjoint of the operator $L$ in Eq.~\ref{l}.
 If $\int e^{-B(x)+rL^-B(x)}dx<\infty$ for
$r$ in some neighborhood of zero, then
$$
\lim_{t\to\infty}\int f L^-B \,dx=\int f_*L^-B\,dx=\int
Lf_*B\,dx=0,$$ which implies
$$
\lim_{t\to\infty}
\dfrac{dH_{G}}{dt}=\lim_{t\to\infty}\dfrac{dH_{c}}{dt}.
$$

\subsubsection{The one dimensional case}\label{ss:1d}

In   a one  dimensional system ($d=1$) the stochastic
differential Eq.~\ref{stochode} becomes
    \be
    \dfrac{dx}{dt} = F(x) +\sigma (x) \xi,
    \label{1dstochode}
    \ee
where  $\xi$ is a (Gaussian   distributed) perturbation
with zero mean and unit variance, and $\sigma (x)$ is the
amplitude of the perturbation. The corresponding
Fokker-Planck equation \ref{fpeqn} is
    \be
    \dfrac{\partial f}{\partial t}
    = - \dfrac {\partial [F(x)f]}{\partial x}
    + \dfrac 12 \dfrac {\partial^2 [\sigma^2(x)f]}{\partial x^2}.
    \label{1dfpeqn}
    \ee

If stationary solutions $f_*(x)$ of  (\ref{1dfpeqn}) exist,
they are  defined by $P^tf_*=f_*$ for all $t$ and given as
the generally unique (up to a multiplicative constant)
solution of
    \be
    - \dfrac {\partial [F(x)f_*]}{\partial x} +
    \dfrac 12 \dfrac {\partial^2 [\sigma^2(x)f_*]}{\partial x^2} = 0.
    \label{fpstatden}
    \ee
The integrable solution is given by
    \be
    f_*(x) = \dfrac {K}{\sigma^2(x)} \exp \left[ \int ^x \dfrac {2F(z)}{\sigma ^2(z)} \,dz \right],
    \label{statden}
    \ee
where $K>0$ is a normalizing constant and the semigroup
$P^t$ is asymptotically stable.

It is known \citep[Section IV]{mackeytyran05} that under
relatively mild conditions there exists a constant $\lambda
>  0$ such that
$$ H_c(P^tf_0|f_*)\ge e^{-2\lambda t}H_c(f_0|f_*).
$$
Specific examples of $\sigma(x)$ and $F(x)$ for which one
can determine the solution $f(t,x)$ of Eq.~\ref{1dfpeqn}
are few. One is that for an Ornstein-Uhlenbeck process
which we consider in our next example.
\begin{example}\label{ex-o-u}  In considering the
Ornstein-Uhlenbeck process, developed in thinking about
perturbations to the velocity of a Brownian particle, we
denote the dependent variable by $v$ so $\sigma(v) \equiv
\sigma$ a constant, and $F(v) = -\gamma v$ with $\gamma
\geq 0$. Now  Eq.~\ref{1dstochode} becomes
    \be
    \dfrac{dv}{dt} = -\gamma v  +\sigma \xi
    \label{o-ueqn}
    \ee
with the Fokker-Planck equation
    \be
    \dfrac{\partial f}{\partial t}
    =  \dfrac {\partial [\gamma v f]}{\partial v}
    + \dfrac {\sigma^2}2 \dfrac {\partial^2 f}{\partial v^2}.
    \label{o-ufpeqn}
    \ee
The unique stationary solution is
    \be
    f_*(v) =  \dfrac {e^{-\gamma v^2/\sigma^2}}{\int_{-\infty}^{+\infty}
    e^{-\gamma v^2/\sigma^2}dv}=  \sqrt{\dfrac {\gamma}{\pi \sigma^2}}e^{-\gamma v^2/\sigma^2}.
    \label{o-ustat}
    \ee

If the initial density $f_0$ is a Gaussian of the form
    \be
    f_0(v) = \dfrac {1}{\bar \sigma \sqrt{2 \pi}} \exp \left \{ -\dfrac {(v - \bar v)^2}{2 \bar \sigma^2} \right
    \}
    \ee
where $\bar \sigma >0$ and $ \bar v \in\realnos$, then
    \be
    P^tf_0 (v) = \dfrac {1}{\sigma_t \sqrt{2 \pi}} \exp \left
    \{ - \dfrac {(v-\bar v(t)) ^2}{2 \sigma_t^2} \right \}
    \ee
wherein
    \be
    \sigma^2_t = \sigma_*^2 + (\bar \sigma^2 -
    \sigma_*^2)e^{-2\gamma t}
    \ee
with $ \sigma_*^2 = \sigma^2/2 \gamma$ and
    \be
    \bar v(t) = \bar v e^{-\gamma t}.
    \ee

The Gibbs' entropy is
 \be
    H_{G}(P^tf_0 ) =  \log \sigma_t \sqrt{2 \pi} + \dfrac 12.
    \ee
Also
    \be
    \int_{-\infty}^{+\infty} P^tf_0 (x) \log f_*(x) dx = -\log
    \sigma_* \sqrt{2 \pi} - \dfrac 12 \dfrac
    {\sigma_t^2}{\sigma_*^2},
    \ee
so
    \begin{equation}
    \begin{split}
    H_c(P^tf_0 |f_*)
     &= \dfrac 12 \log \left [\dfrac  {\sigma_t^2}{\sigma_*^2}
    \right ]  + \dfrac 12 \left [ 1 - \dfrac
    {\sigma_t^2}{\sigma_*^2}\right ] \\
    &=\dfrac 12 \log \left \{ 1 + e ^{-2\gamma t} \left [ \dfrac  {\bar \sigma
    ^2}{\sigma_*^2}- 1
    \right ] \right \} \\ &\quad - \dfrac 12 e ^{-2 \gamma t} \left [ \dfrac
    {\bar \sigma^2}{\sigma_*^2} - 1 \right ].
    \end{split}
    \end{equation}

We may examine how the two  different types of entropy behave.
First, we may show that $\dot H_c(P^tf_0 |f_*) \geq 0$ with
    $$
    \left ( \dfrac{dH_c(P^tf_0 |f_*)}{dt} \right ) _{t = 0} =
      \dfrac{\gamma(R-1)^2}{R} > 0
    $$
and $H_c(P^tf_0 |f_*)$ increasing in a monotone fashion, with
    $$
    \lim_{t \to \infty} \dfrac{dH_c(P^tf_0 |f_*)}{dt} =
    0,
    $$
wherein $R \equiv \bar \sigma^2/\sigma_*^2$.

This  is not
the case with the   Gibbs' entropy,  for
    \be \dfrac {dH_{G}(P^tf_0 )}{dt}  \left \{
    \begin{array}{lll}
    &>& 0  \qquad \mbox{for} \qquad \bar \sigma^2 < \sigma_*^2 \\
    &=& 0  \qquad \mbox{for}  \qquad \bar \sigma^2 = \sigma_*^2\\
    &<& 0   \qquad \mbox{for}  \qquad \bar \sigma^2 > \sigma_*^2
    ,
    \end{array} \right.  \ee
 implying that the evolution  of the Gibbs' entropy in time is
a function of the statistical properties ($\bar \sigma ^2$)
of the initial ensemble.  All of these conclusions
concerning the dynamics of $H_G(P^tf_0 )$ are implicit in
the work of \citet{bag02} but not explicitly stated.

\end{example}

Similar effects can be observed for the Rayleigh process considered
in \citep[Section IV]{mackeytyran05}.

\subsubsection{Multidimensional Ornstein-Uhlenbeck process}\label{ss:mo-up}

Consider the multidimensional Ornstein-Uhlenbeck process
    \be
    \dfrac{dx}{dt} = Fx  +\Sigma \xi,
    \label{linear}
    \ee
where $F$ is a $d\times d$ matrix, $\Sigma$ is a $d\times
d$ matrix an $\xi$ is $d$ dimensional vector. The formal
solution to Eq.~\ref{linear} is given by
    \be
    x(t)=e^{tF}x(0)+\int_0^t e^{(t-s)F}\Sigma \, dw(t),
    \label{slinear}
    \ee
where $e^{tF}=\sum_{n=0}^\infty \frac{t^n}{n!}F^n$ is the
fundamental solution to $\dot{X}(t)=FX(t)$ with $X(0)=I$, and $w(t)$
is the standard $d$-dimensional Wiener process. From the properties
of stochastic integrals it follows that
$$
\eta(t)=\int_0^t e^{(t-s)F}\Sigma\, dw(t)
$$
has mean $0$ and covariance
\begin{equation}
R(t)=E\eta(t)\eta(t)^T=\int_{0}^t e^{sF}\Sigma\Sigma^T
e^{sF^T} ds,
\end{equation} where $F^T$ is the
transpose of the matrix $F$. The matrix $R(t)$ is
nonnegative definite but not necessarily positive definite.
We follow the presentation of \cite{zakaisnyders70} and
\cite{erickson71}. For each $t>0$ the matrix $R(t)$ has
constant rank equal to the dimension of the space
$$
[F,\Sigma]:=\{F^{l-1}\Sigma \epsilon_j: l,j=1,\ldots,d,
\epsilon_j=(\delta_{j1},\ldots,\delta_{jp})^T\}.
$$
If $l=\rank R(t)$ then $d-l$ coordinates of the process
$\eta(t)$ are equal to $0$ and the remaining $l$
coordinates constitute an $l$-dimensional Gaussian process.
Thus if $l<d$ there is no stationary density. If $\rank
R(t)=d$ then the transition probability function of $x(t)$
is given by the Gaussian density
    \be
    k(t,x,x_0)=\frac{\exp\{-\frac{1}{
    2}(x-e^{tF}x_0)^TR(t)^{-1}(x-e^{tF}x_0)\}}{\sqrt{(2\pi)^{d}\det
    R(t)}},
    \label{e:tdmo-up}
    \ee
where $R(t)^{-1}$ is the inverse matrix of $R(t)$. An
invariant density $f_*$ exists if and only if all
eigenvalues of $F$ have negative real parts, and in this
case the unique stationary density $f_*$ has the form
    \be
    f_*(x)=\frac{1}{\sqrt{(2\pi)^{d}\det
    R_{*}}}\exp\left \{-\frac{1}{2}x^T R_{*}^{-1}x\right \},
    \label{e:sdmo-up}
    \ee
where $R_{*}$ is a positive definite matrix given by
    $$
    R_{*}=\int_0^\infty e^{sF}\Sigma\Sigma^T e^{sF^T} ds,
    $$
    and is a unique symmetric matrix satisfying
    \be
    F R_*+R_*F^T=-\Sigma\Sigma^T.\label{e:Leq}
    \ee
We conclude that if $[F,\Sigma]$ contains $d$ linearly
independent vectors and all eigenvalues of $F$ have
negative real parts, then the corresponding semigroup of
Markov operators is asymptotically stable. From
Theorem~\ref{t:entropyconv} it follows that
$$\lim_{t \to \infty}H_c(P^tf_0 |f_*) = 0$$ and from
Theorem~\ref{th-convg} that
$$\lim_{t \to \infty}H_G(P^tf_0 ) = H_G(f_*)$$
for all $f_0$ with $H_c(f_0|f_*)>-\infty$.

Now let $f_0$ be a Gaussian density of the form
    \be
    f_0(x)=\frac{1}{\sqrt{(2\pi)^{d}\det
    Q_0}}\exp\left \{-\frac{1}{2}x^T Q_0^{-1}x\right \},  \label{e:gaussini}
    \ee
where $Q_0$ is a positive definite symmetric matrix. From
Eq.~\ref{slinear} it follows that $x(t)$ is Gaussian with
zeroth mean vector and  the following covariance matrix
    \be
   Q(t)=e^{tF}Q_0 e^{tF^T}+ R(t).
   \label{e:o-umdm}
  \ee
Hence the density of $x(t)$ is given by
    \be
    P^tf_0(x)=\frac{1}{\sqrt{(2\pi)^{d}\det
    Q(t)}}\exp\left \{-\frac{1}{2}x^T Q(t)^{-1}x\right \}.
    \label{e:gaussinif}
    \ee
Since $\int P^tf_0 (x) x^T Q(t)^{-1}x\, dx=d$, the Gibbs'
entropy of $P^tf_0 $  is
    \be
    H_G(P^tf_0 )=\dfrac 12\log (2\pi)^{d}\det
    Q(t) + \dfrac{d}{2}.\label{e:generalg}
    \ee
By Eq.~\ref{d:con-bg-ent1} and the formula
$$
\int P^tf_0 (x)x^TR_*^{-1} x\, dx=\Tr (R_*^{-1}Q(t))
$$
we obtain the conditional entropy
\begin{equation}\begin{split} H_c(P^tf_0 |f_*)&=H_G(P^tf_0 )-\dfrac 12 \log(2\pi)^{d}\det
R_*\\ &\quad -\dfrac 12
\Tr(R_*^{-1}Q(t))\label{f:cemou1}
    \end{split}
    \end{equation}
for all $t\ge 0$ and every $f_0$ of the form given by
Eq.~\ref{e:gaussini}. Formula~\ref{e:generalg} remains
valid when we start with a Gaussian density $f_0$ with non
zero mean vector but then in the formula for the
conditional entropy one additional term appears, see
\citep[Section IV]{mackeytyran05}.

\begin{example} {Noisy harmonic oscillator.}\label{sss:ho}

  Consider the second order system
    \be
    m\dfrac{d^2y}{dt^2}+ \gamma \dfrac{dy}{dt}+\omega^2 y=\sigma \xi
    \label{e:b-oscn}
    \ee
with constant positive coefficients $m$, $\gamma$ and
$\sigma$. Introduce the velocity $v=\dfrac{dy}{dt}$ as a
new variable. Then Eq.~\ref{e:b-oscn} is equivalent to the
system
    \begin{subequations}
    \begin{align}
   \label{e:b-oscn1} \dfrac{dy}{dt} &= v \\
    m \dfrac{dv}{dt} &= -\gamma v-\omega^2 y +\sigma \xi,\label{e:b-oscn2}
    \end{align}
    \end{subequations}
and the corresponding Fokker-Planck equation is
    $$
    \dfrac{\partial f}{\partial t}
    = - \dfrac {\partial [v f]}{\partial y}+ \dfrac {1}{m} \dfrac {\partial [(\gamma v +\omega^2 y )f]}{\partial v}
    + \dfrac {\sigma^2}{2 m^2} \dfrac {\partial^2 f}{\partial v^2}.
       $$
We can assume in what follows that $m=1$, as introducing
the constants $\tilde{\gamma}=\gamma/m$,
$\tilde{\omega^2}=\omega^2/m$ and
$\tilde\sigma^2=\sigma^2/m^2$ leads to
$$
    \dfrac{\partial f}{\partial t}
    = - \dfrac {\partial [v f]}{\partial y}+ \dfrac {\partial [(\tilde{\gamma} v +\tilde{\omega^2} y )f]}{\partial v}
    + \dfrac {\tilde{\sigma^2}}{2} \dfrac {\partial^2 f}{\partial
v^2}.
       $$

The results of Section~\ref{ss:mo-up} in the two
dimensional setting apply with $x=(y,v)^T$,
$$
F=\left(\begin{array}{cc} 0 & 1\\
-\omega^2& -\gamma
\end{array}\right),
\quad\mbox{and}\quad \Sigma=\left(\begin{array}{cc}0 & 0\\
0 & \sigma
\end{array}\right).
$$
Since
$$
[F,\Sigma]=\left\{ \left(
\begin{array}{c}
 0 \\
 0 \\
\end{array}
\right), \left(
\begin{array}{c}
  0 \\
  \sigma \\
\end{array}%
\right),
\sigma \left(%
\begin{array}{c}
  1 \\
  -\gamma \\
\end{array}%
\right) \right\},
$$
the transition density function is given by
Eq.~\ref{e:tdmo-up}. The eigenvalues of $F$ are equal to
\begin{subequations}\label{e:eigenvalues}
\begin{align}
 \lambda_1&=\frac{-\gamma+\sqrt{\gamma^2-4
\omega^2}}{2},\\
\lambda_2&=\frac{-\gamma-\sqrt{\gamma^2-4 \omega^2}}{2},
\end{align}
\end{subequations}
and are either negative real numbers when $\gamma^2\ge
4\omega^2$ or complex numbers with negative real parts when
$\gamma^2< 4 \omega^2$.  Thus the stationary density is
given by Eq.~\ref{e:sdmo-up}. As is easily seen $R_*$,
being a solution to Eq.~\ref{e:Leq}, is given by
$$
R_{*}=\dfrac{\sigma^2}{2\gamma\omega^2}\left(\begin{array}{cc}
1& 0 \\
0 & \omega^2\end{array}\right).
$$
The inverse of the matrix $R_*$ is
$$
R_{*}^{-1}=\dfrac{2\gamma}{\sigma^2} \left(\begin{array}{cc}
\omega^2 & 0 \\
0 & 1\end{array}\right)
$$
and the unique stationary density becomes
$$
f_*(y,v)=  \dfrac {\gamma \omega }{\pi \sigma ^2}
e^{-\dfrac {\gamma }{\sigma ^2}[\omega ^2 y^2+ v^2 ]}.
$$

If  the initial density $f_0$ is the Gaussian
$$
f_0(y,v)=\dfrac{1}{2\pi \bar\sigma_1\bar\sigma_2 }\exp
\left\{-\dfrac{y^2}{2\bar\sigma_1^2}-\dfrac{v^2}{2\bar\sigma_2^2}\right\},
$$
where $\bar\sigma_1>0,\bar\sigma_2>0$, then $P^tf_0 $ is as
in Eq.~\ref{e:gaussinif} with
$$
Q(t)=e^{tF}Q_0e^{tF^T}+R(t),
$$
where
\begin{equation}\label{e:matrixQ0}
    Q_0=\left(\begin{array}{cc} \bar \sigma_1^2 & 0\\
0 & \bar \sigma_2^2\end{array}\right)
\end{equation}
and
\begin{equation}\label{e:matrixR}
R(t)=\int_0^te^{sF}\left(\begin{array}{cc} 0 & 0\\
0 & \sigma^2\end{array}\right)e^{sF^T}ds.
\end{equation}
The formula for the covariance matrix $R(t)$ is given by
\citet[pp. 27-30]{chandra43}. The Gibbs' entropy is
\begin{equation}\label{e:hformulagibbs}
H_G(P^tf_0 )=1+\log (2\pi)+\dfrac 12 \log\det
    Q(t)
\end{equation}
and  the conditional entropy is
\begin{equation}\label{e:hformulacond}
\begin{split}
H_c(P^tf_0 |f_*)& =1+\dfrac 12 \log\det Q(t)-\dfrac 12\log \det R_*\\
& \quad -\dfrac 12\Tr (R_*^{-1}Q(t)).
\end{split}
\end{equation}
We are going to show that the Gibbs entropy need  not be a monotonic
function of time, so we need to calculate $\det Q(t)$ to have the
analytic formula for the  Gibbs entropy. The calculations depend on the nature
of eigenvalues $\lambda_1$ and $\lambda_2$ in
Eq.~\ref{e:eigenvalues}, so we must distinguish between  three cases:
(i) $\lambda_1,\lambda_2\in \realnos$ with $\lambda_1\neq\lambda_2$,
(ii) $\lambda_1,\lambda_2\in \realnos$ with $\lambda_1=\lambda_2$,
and (iii) $\lambda_1,\lambda_2$ are complex.

In what follows we use the following notation
\begin{subequations}\label{e:nalpha}
\begin{align}
\label{e:nsigma}
\sigma_*&=\dfrac{\sigma^2}{2\gamma\omega^2},\\
\label{e:nalpha1}  \alpha_1 &= \bar\sigma_1^2-\sigma_*, \\
\label{e:nalpha2}  \alpha_2 &= \bar\sigma_2^2-\omega^2\sigma_*.
\end{align}
\end{subequations}
Observe that $\alpha_1\alpha_2=\det(Q_0-R_*)$ and
$\sigma_*^2\omega^2=\det R_*$.

(i) Let us first consider the {\bf overdamped case}
$$
\gamma^2>4\omega^2,
$$
so the eigenvalues in Eq.~\ref{e:eigenvalues} are real and
$\lambda_1\neq \lambda_2$. Define, for $t\ge 0$,
    \begin{subequations}\label{e:hc}\begin{align}\label{e:hc1}
    c_1(t)&=\frac{\lambda_2 e^{\lambda_1 t}-\lambda_1 e^{\lambda_2
    t}}{\lambda_2-\lambda_1},\\
    \label{e:hc2}
    c_2(t)&=\frac{e^{\lambda_2 t}-e^{\lambda_1
    t}}{\lambda_2-\lambda_1}.
    \end{align}
    \end{subequations}
Then
$$
e^{tF}=\left(\begin{array}{cc} c_1(t)& c_2(t)\\
c'_1(t) & c'_2(t)\end{array}\right)
$$
and the covariance matrix $R(t)$ is given by
$$
R(t)=R_*
-\dfrac{\sigma^2}{2\gamma\omega^2} \left(\begin{array}{cc}
c_1^2+\omega^2 c_2^2 &-\gamma\omega^2c_2^2 \\
-\gamma\omega^2c_2^2 & (c_1')^2+\omega^2(c'_2)^2\end{array}\right),
$$
where we suppressed the dependence of $c_1$ and $c_2$ on
$t$. Accordingly, for $Q_0$ as in Eq.~\ref{e:matrixQ0} we
have
$$
e^{tF}Q_0e^{tF^T}=
\left(\begin{array}{cc}
c_1^2 \bar\sigma_1^2+c_2^2 \bar\sigma_2^2
& c_1c_1' \bar\sigma_1^2+c_2c_2'\bar\sigma_2^2\\
c_1c_1'\bar\sigma_1^2+c_2c_2'\bar\sigma_2^2 &
(c'_1)^2\bar\sigma_1^2+(c'_2)^2\bar\sigma_2^2
\end{array}\right).
$$
 From Eq.~\ref{e:hc} it follows that
$$
c_1c_1'+\omega^2c_2c_2'=-\gamma \omega^2 c_2^2.
$$
Combining the three preceding equations and introducing the values
of $\sigma_*$, $\alpha_1$, and $\alpha_2$ from
Eq.~\ref{e:nalpha}, 
we obtain for the matrix
$Q(t)$ the formula
$$
Q(t)=
\left(\begin{array}{cc}
c_1^2\alpha_1+c_2^2\alpha_2 +\sigma_*& c_1c_1'\alpha_1+c_2c_2'\alpha_2\\
c_1c_1'\alpha_1+c_2c_2'\alpha_2 & (c'_1 )^2\alpha_1+
(c'_2)^2\alpha_2+\omega^2\sigma_*
\end{array}\right).
$$
Hence
\begin{equation}\nonumber\begin{split}
\det Q(t)&=\omega^2\sigma_*^2+\alpha_1\alpha_2
(c_1c_2'-c_1'c_2)^2\\
& \quad +\sigma_*
\left(\left(\omega^2c_1^2+(c_1')^2\right)\alpha_1+\left(\omega^2
c_2^2+(c'_2)^2\right)\alpha_2\right).
\end{split}
\end{equation}
Making use of Eq.~\ref{e:hc} together with the relations
$\lambda_1\lambda_2=\omega^2$ and
$\lambda_1+\lambda_2=-\gamma$, we arrive at
\begin{equation}\nonumber
\begin{split}
\det Q(t)&=\omega^2\sigma_*^2+ \alpha_1\alpha_2 e^{-2\gamma t}\\
& \quad -\dfrac{\sigma_*}{
(\lambda_1-\lambda_2)^2}\left(\gamma\lambda_1(\lambda_2^2\alpha_1+\alpha_2)e^{2\lambda_1t}\right.\\
& \quad\left. +4\omega^2 (\omega^2\alpha_1+\alpha_2)e^{-\gamma
t}+\gamma\lambda_2(\lambda_1^2\alpha_1+\alpha_2)e^{2\lambda_2t}
\right).
\end{split}
\end{equation}
Consequently, after some algebra we obtain
 \be
 \dfrac{d H(P^tf_0)}{dt}=- \dfrac{\gamma
\left(\alpha_1\alpha_2e^{-2\gamma
t}+\sigma_*\left(c_2^2\omega^4\alpha_1+(c_2')^2\alpha_2\right)\right)}{\det
Q(t)}\label{e:ohdergibbs}
 \ee
 and
 \be \left(\dfrac
{dH_{G}(P^tf_0 )}{dt}\right)_{t=0}=- \dfrac{\gamma\alpha_2}{\bar\sigma_2^2}
\left \{
    \begin{array}{lll}
     <& 0  \quad \mbox{for} \quad \alpha_2> 0, \\
>& 0  \quad \mbox{for} \quad \alpha_2< 0.\\
 \end{array} \right. \label{e:ohdergibbs0} \ee
 Since $\gamma>0$ and $\det Q(t)>0$, the sign of the derivative of
 $H_G(P^tf_0 )$ is completely determined by the remaining parts
 and depends on the sign of $\alpha_1$ and $\alpha_2$ and their mutual relations.
In the case of $\alpha_1\alpha_2=0$ we conclude from
Eqs.~\ref{e:ohdergibbs} and~\ref{e:nalpha} that
 \be \label{e:hozero}\dfrac {dH_{G}(P^tf_0 )}{dt}  \left \{
    \begin{array}{lll}
=& 0   \quad \mbox{for}  \quad  \,\bar\sigma_1^2=\sigma_*, \,\,\bar\sigma_2^2=\omega^2\sigma_*,\\
>& 0  \quad \mbox{for} \quad \begin{array}{ll}
 \bar\sigma_1^2=\sigma_*,& \bar\sigma_2^2<\omega^2\sigma_*, \\
 \bar\sigma_1^2<\sigma_*,&\bar\sigma_2^2=\omega^2\sigma_*,
\end{array}\\
<& 0  \quad \mbox{for} \quad \begin{array}{ll} \bar\sigma_1^2=\sigma_*, &\bar\sigma_2^2>\omega^2\sigma_*,\\
 \bar\sigma_1^2>\sigma_*, & \bar\sigma_2^2=\omega^2\sigma_*\end{array}
    \end{array} \right.  \ee
    for all $t\ge 0$.
Now assume that $\alpha_1\alpha_2\neq 0$. It also follows directly
from Eq.~\ref{e:ohdergibbs} that
 \be \dfrac {dH_{G}(P^tf_0 )}{dt}
     < 0  \quad \mbox{for} \quad \bar\sigma_1^2>\sigma_*, \bar\sigma_2^2>\omega^2\sigma_*. \label{e:homon}  \ee
This behavior is  illustrated in Fig.~\ref{f:ohmon}.
\begin{figure}[ht]
\includegraphics[scale=0.5]{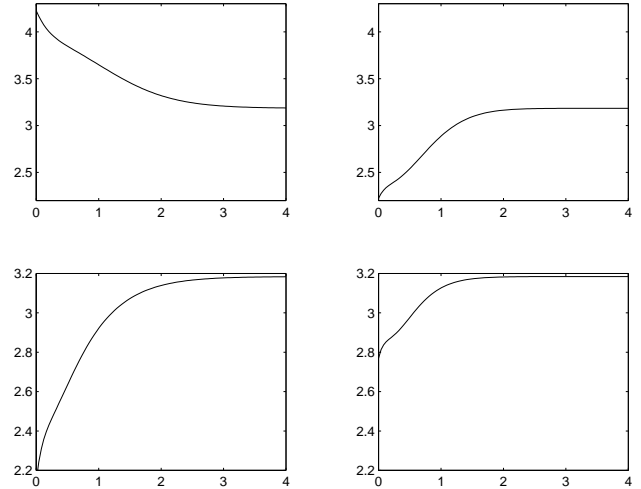}
\caption{Entropy behavior for the overdamped noisy harmonic oscillator.
The left hand  panels show plots of $H_G(P^tf_0 )$ as a function of time
as given by Eq.~\ref{e:hformulagibbs} and the right hand panels show
$H_c(P^tf_0 |f_*)$ (as in Eq.~\ref{e:hformulacond})
plus $H_G(f_*)$, i.e. $H_c(P^tf_0 |f_*) + H_G(f_*)$ as a function of time.
The parameters used were $m=1$, $\gamma=3$, $\omega^2=2$,
 and $\sigma_*=1$. Upper panels correspond to the range of parameters as
in Eq.~\ref{e:homon} with specific values $\bar\sigma_1=2$,
$\bar\sigma_2=2$, while the lower panels correspond to parameters as in
Eq.~\ref{e:homon1} with $\bar\sigma_1=0.5$, $\bar\sigma_2=1$.}
\label{f:ohmon}
\end{figure}

 To  study the remaining cases we rewrite
 Eq.~\ref{e:ohdergibbs} in the form
 \be \dfrac {dH_{G}(P^tf_0 )}{dt}= \dfrac{\gamma}{\det
Q(t)}e^{-2\gamma t}h_1(t),\ee where
   \begin{equation}
    \begin{split}
    h_1(t)&=-\alpha_1\alpha_2 -\sigma_*
\left(\lambda_1\beta_1e^{-2\lambda_2t}+\lambda_2\beta_2e^{-2\lambda_1t}\right)\\
& \quad -2\sigma_*\dfrac{\omega^2}{\gamma}
(\beta_1+\beta_2)e^{-\gamma t}
 \end{split}
 \end{equation}
and
\begin{subequations}\label{e:ohbeta}
\begin{align}
 \label{e:ohbeta1} \beta_1 &= \dfrac{\lambda_1(\lambda_2^2\alpha_1+\alpha_2)}{(\lambda_1-\lambda_2)^2}, \\
 \label{e:ohbeta2} \beta_2 &= \dfrac{\lambda_2(\lambda_1^2\alpha_1+\alpha_2)}{(\lambda_1-\lambda_2)^2}.
 \end{align}
\end{subequations}
 Since
$\lambda_1\lambda_2=\omega^2$, we obtain
$$
h_1'(t)=2\omega^2\sigma_*
\left(\beta_1e^{-2\lambda_2t}-(\beta_1+\beta_2)e^{-\gamma
t}+\beta_2e^{-2\lambda_1t}\right),
$$
which leads to \be\nonumber h_1'(t)=2\omega^2\sigma_*
e^{-2\lambda_1t}\left(e^{(\lambda_1-\lambda_2)t}-1\right)
\left(\beta_1e^{(\lambda_1-\lambda_2)t}- \beta_2\right).
\ee For $t_*>0$ such that
$\beta_1e^{(\lambda_1-\lambda_2)t_*}= \beta_2$ we have \be
h_1(t_*)= -\alpha_1\left(\alpha_2+\omega^2\sigma_*
\left(\dfrac{\beta_2}{\beta_1}\right)^{\gamma/(\lambda_1-\lambda_2)}\right).\label{e:ohh1ex}
\ee Returning to formulae~\ref{e:ohbeta}, we note that
$$
\dfrac{\beta_2}{\beta_1}=
1+\dfrac{(\lambda_1-\lambda_2)(\omega^2\alpha_1-\alpha_2)}{\lambda_1(\lambda_2^2\alpha_1+\alpha_2)}.
$$

We can now continue to  study the of behavior of
$H_G(P^tf_0 )$. First, we consider the case of $\alpha_1<0$
and $\alpha_2<0$. If $\omega^2\alpha_1\ge\alpha_2$ then
$h_1(t)\ge h_1(0)$ and $h_1(0)>0$ by
Eq.~\ref{e:ohdergibbs0}. Now if $\omega^2\alpha_1<\alpha_2$
then $h_1(t)\ge h_1(t_*)$ and from Eq.~\ref{e:ohh1ex} it
follows that $h_1(t_*)>0$. Consequently, we obtain
 \be \dfrac {dH_{G}(P^tf_0 )}{dt}
     > 0  \quad \mbox{for} \quad \bar\sigma_1^2<\sigma_*, \bar\sigma_2^2<\omega^2\sigma_*. \label{e:homon1}  \ee

Consider now the case of $\alpha_1>0$ and $\alpha_2<0$. Then we
know that $h_1(0)>0$. 
Now if $\lambda_2^2\alpha_1+\alpha_2\ge 0$ 
then $\beta_1\le 0$ and $\beta_1\le\beta_2$. Thus $h_1$ is
decreasing and diverges to $-\infty$ as $t\to\infty$. Consequently,
if $\lambda_2^2\alpha_1\ge -\alpha_2>0$, or equivalently
\begin{equation}\label{e:ohpar1}
  \lambda_2^2\bar\sigma_1^2+\bar\sigma_2^2+\gamma\lambda_2\sigma_*\ge 0\quad\mbox{and}\quad
  \bar\sigma_2^2<\omega^2\sigma_*,
\end{equation} then there is $t_0>0$ such that \be \dfrac {dH_{G}(P^tf_0 )}{dt}\left
\{
    \begin{array}{ll}
>& 0   \qquad \mbox{for} \qquad t< t_0, \\
<& 0   \qquad \mbox{for} \qquad t> t_0.
\end{array} \right.\label{e:hopn} \ee
\begin{figure}[htb]
\includegraphics[scale=0.5]{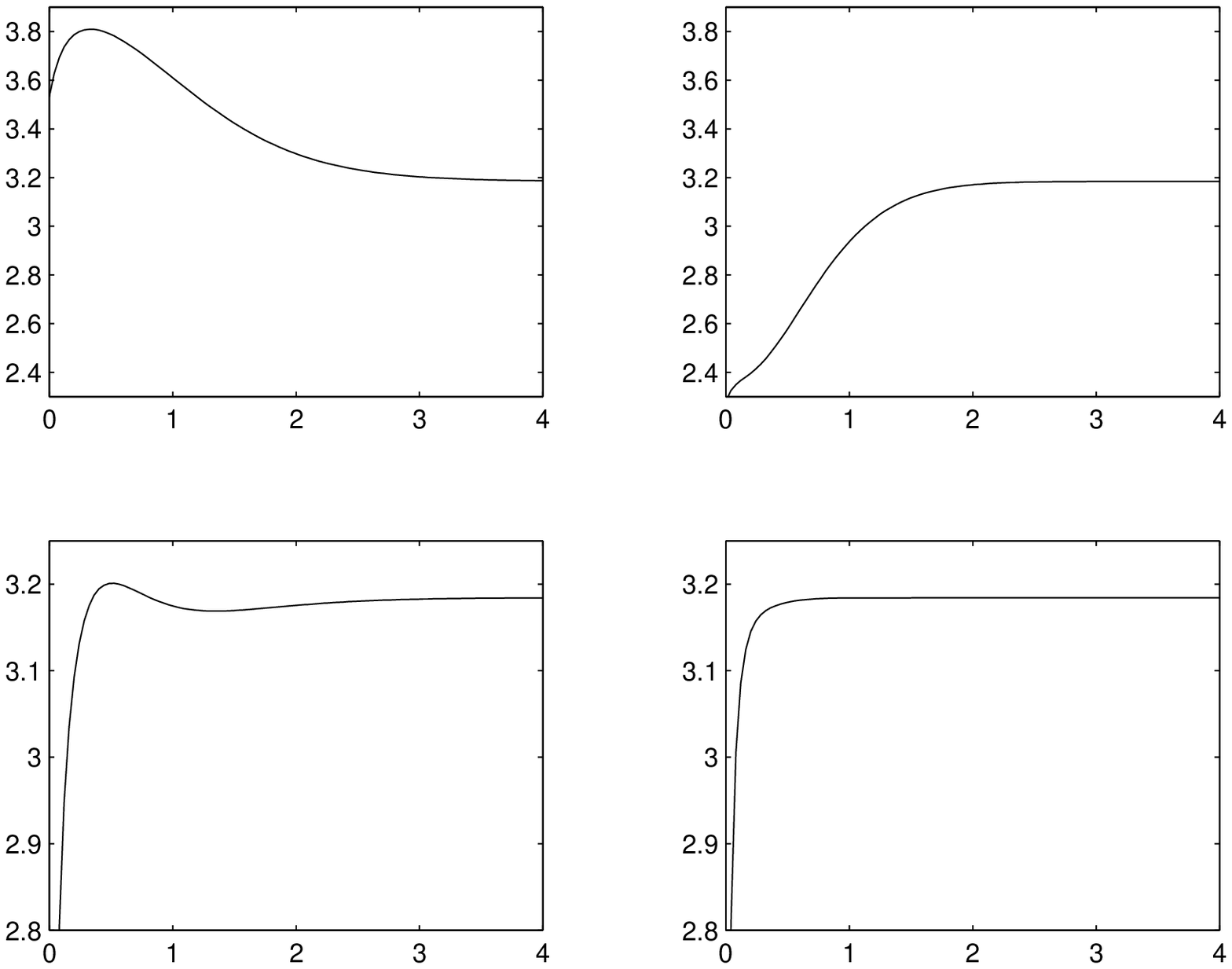}
\caption{Overdamped noisy harmonic oscillator. Plots and
parameters as in Fig.~\ref{f:ohmon}, but now upper panels
are for the range of parameters as in Eq.~\ref{e:ohpar1}
with $\bar\sigma_1=2$, $\bar\sigma_2=1$, and lower as in
Eq.~\ref{e:ohpar2} with $\bar\sigma_1=1.1$,
$\bar\sigma_2=0.1$.} \label{f:ohmon1}\end{figure} If
$\lambda_2^2\alpha_1+\alpha_2<0$ then $\beta_1>0$ and
$\beta_2/\beta_1>1$. From Eq.~\ref{e:ohh1ex} it follows
that $h_1(t_*)<0$. Thus $h_1$ starting from a positive
value at $0$ decreases to a negative value at $t_*$ and
then increases and diverges to $\infty$. Hence we conclude
that, if $0<\lambda_2^2\alpha_1< -\alpha_2$, or
equivalently
\begin{equation}\label{e:ohpar2}
  \lambda_2^2\bar\sigma_1^2+\bar\sigma_2^2+\gamma\lambda_2\sigma_*<0\quad\mbox{and}\quad
  \bar\sigma_1^2>\sigma_*,
\end{equation} then there are $t_1,t_2>0$ such
that \be \dfrac {dH_{G}(P^tf_0 )}{dt}\left \{
    \begin{array}{ll}
>& 0   \qquad \mbox{for} \qquad 0<t< t_1, \\
<& 0   \qquad \mbox{for} \qquad t_1<t<t_2,\\
>& 0   \qquad \mbox{for} \qquad t> t_2. \\
\end{array} \right.\label{e:hopnp} \ee
These behaviors are illustrated in Fig.~\ref{f:ohmon1}.

A symmetric behavior is observed  when $\alpha_1<0$ and
$\alpha_2>0$, and graphically shown in Fig.~\ref{f:ohmon2}.
\begin{figure}[htb]
\includegraphics[scale=0.5]{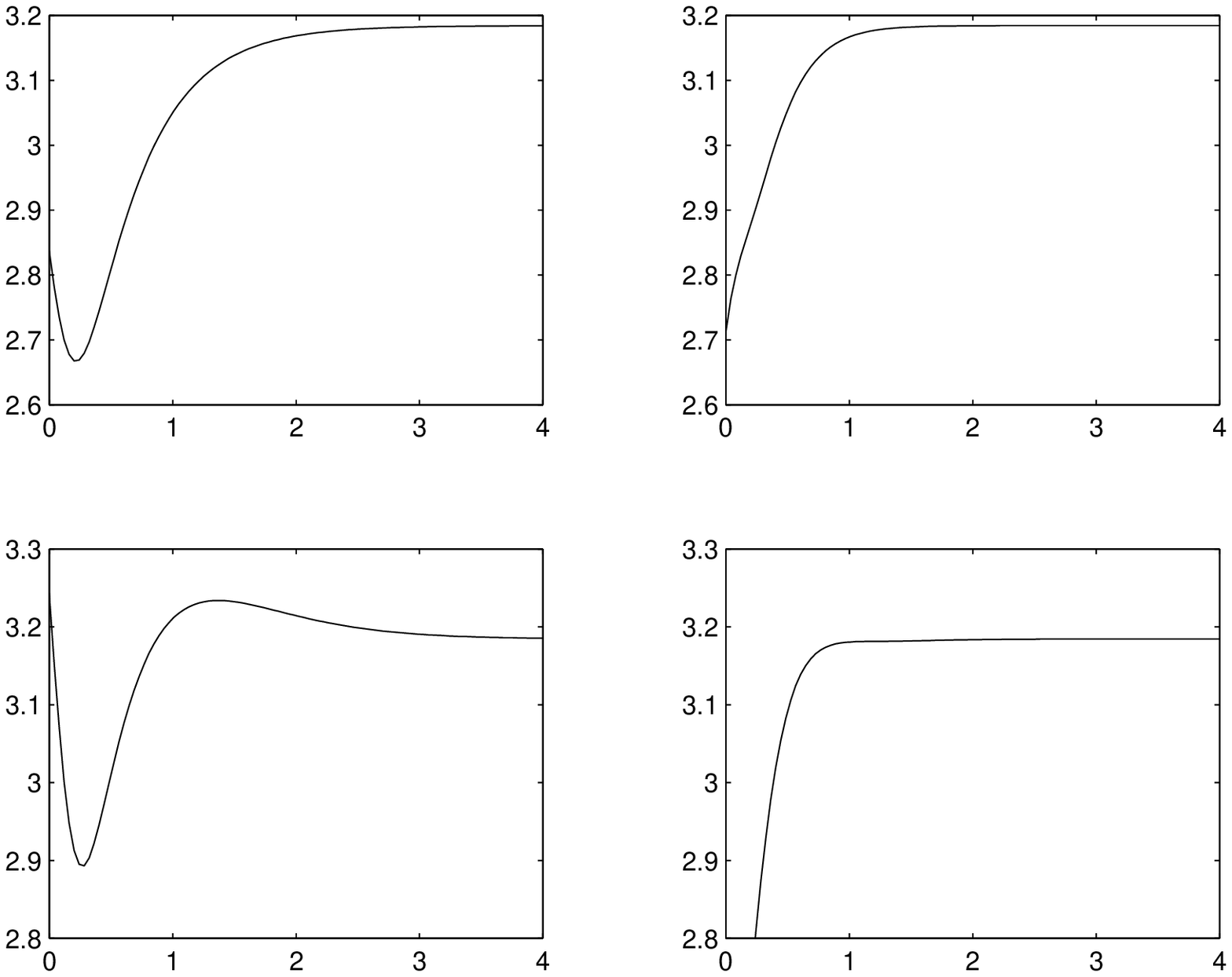} \caption{Overdamped noisy harmonic oscillator. Plots and paramenters as in Fig.~\ref{f:ohmon}, but the  upper panels are for
the range of parameters as in Eq.~\ref{e:ohpar3} with
$\bar\sigma_1=0.5$, $\bar\sigma_2=2$, and lower as in
Eq.~\ref{e:ohpar4} $\bar\sigma_1=0.5$, $\bar\sigma_2=3$.}
\label{f:ohmon2}\end{figure} We then have $h_1(0)<0$ and a
similar analysis leads to the following conclusions. If
$\lambda_2^2\alpha_1\le-\alpha_2<0$,  or equivalently
\begin{equation}\label{e:ohpar3}
  \lambda_2^2\bar\sigma_1^2+\bar\sigma_2^2+\gamma\lambda_2\sigma_*\le 0\quad\mbox{and}\quad
  \bar\sigma_2^2>\omega^2\sigma_*,
\end{equation}
then there is $t_0>0$ such that \be \dfrac {dH_{G}(P^tf_0 )}{dt}\left
\{
    \begin{array}{ll}
<& 0   \qquad \mbox{for} \qquad t< t_0, \\
>& 0   \qquad \mbox{for} \qquad t> t_0
\end{array} \right.\label{e:honp} \ee
and if $0>\lambda_2^2\alpha_1>-\alpha_2$, or equivalently
\begin{equation}\label{e:ohpar4}
  \lambda_2^2\bar\sigma_1^2+\bar\sigma_2^2+\gamma\lambda_2\sigma_*>0\quad\mbox{and}\quad
  \bar\sigma_1^2<\sigma_*,
\end{equation} then  there are
$t_1,t_2>0$ such that \be \dfrac {dH_{G}(P^tf_0 )}{dt}\left \{
    \begin{array}{lll}
<& 0   \qquad \mbox{for} \qquad 0<t< t_1, \\
>& 0   \qquad \mbox{for} \qquad t_1<t<t_2,\\
<& 0   \qquad \mbox{for} \qquad t> t_2. \\
\end{array} \right. \label{e:honpn}\ee


(ii) Let us now consider  the {\bf critical damping} situation when
$$
\gamma^2=4\omega^2,
$$
so that $\lambda_1=\lambda_2$, 
and set
$$
\lambda=-\frac{\gamma}{2}.
$$
In this case we have
$$
F=\left(%
\begin{array}{cc}
  0 & 1 \\
  -\lambda^2 & 2\lambda \\
\end{array}%
\right) \quad\mbox{and}\quad
e^{tF}=e^{\lambda t}\left(%
\begin{array}{cc}
  1-\lambda t & t \\
  -\lambda^2 t & 1+\lambda t \\
\end{array}%
\right),
$$
so that the corresponding covariance matrix $R(t)$ is given
by
$$
R(t)=R_*+\dfrac{\sigma^2 e^{2\lambda
  t}}{4\lambda^3}
\left(%
\begin{array}{cc}
 (1-\lambda t)^2+\lambda^2 t^2
  & 2\lambda^3 t^2 \\
   2\lambda^3 t^2
  & (\lambda+\lambda^2 t)^2 +\lambda^4 t^2\\
\end{array}%
\right).
$$
We also have
\begin{widetext}
$$
e^{tF}Q_0e^{tF^T}=e^{2\lambda t}\left(%
\begin{array}{cc}
  \bar\sigma_1^2(1-\lambda t)^2 +\bar\sigma_2^2t^2 & -\bar\sigma_1^2\lambda^2t(1-\lambda t)+\sigma_2^2t(1+\lambda t) \\
 -\bar\sigma_1^2\lambda^2t(1-\lambda t)+\sigma_2^2t(1+\lambda t) & \bar\sigma_1^2\lambda^4t^2+\bar\sigma_2^2(1+\lambda t)^2 \\
\end{array}%
\right).
$$
\end{widetext}
Note that now $\sigma_*=- \dfrac{\sigma^2}{4\lambda^3}$ and
$\omega^2=\lambda^2$. Thus
\begin{widetext}
$$
Q(t)=\left(%
\begin{array}{cc}
  e^{2\lambda t}(\alpha_1(1-\lambda t)^2 +\alpha_2t^2)+\sigma_* & e^{2\lambda t}\left(-\lambda^2 t^2(1-\lambda t)\alpha_1+t(1+\lambda t)\alpha_2\right) \\
 e^{2\lambda t}\left(-\lambda^2 t^2(1-\lambda t)\alpha_1+t(1+\lambda t)\alpha_2\right) & e^{2\lambda t}(\alpha_1\lambda^4t^2+\alpha_2(1+\lambda t)^2)+\lambda^2\sigma_* \\
\end{array}%
\right),
$$
\end{widetext}
where $\alpha_1$ and $\alpha_2$ are given by Eq.~\ref{e:nalpha}.
Hence
\begin{equation}\nonumber
\begin{split}
\det Q(t)&=\lambda^2\sigma_*^2+\alpha_1\alpha_2e^{4\lambda
t}+\sigma_*e^{2\lambda t}(\alpha_1\lambda^2((1-\lambda t)^2
\\
&\quad +\lambda^2t^2)+\alpha_2((1+\lambda t)^2+\lambda^2t^2))
\end{split}
\end{equation}
and after some algebra we obtain 
\begin{equation}
\begin{split}
\dfrac{d H_G(P^tf_0 )}{dt}&=\dfrac{2\lambda}{\det Q(t)} e^{4\lambda
t}(\alpha_1\alpha_2+\sigma_*\alpha_1\lambda^4 t^2e^{-2\lambda t}
\\&\quad +\sigma_*\alpha_2(1+\lambda t)^2e^{-2\lambda t}).
\end{split}\label{e:chdergibbs}
\end{equation}
Since $\lambda=-\gamma/2$, Eq.~\ref{e:ohdergibbs0} remains
valid. Now the analysis and conclusions are similar to the
overdamped case. First, observe that from
Eq.~\ref{e:chdergibbs} follow Eq.~\ref{e:hozero} in the
case of $\alpha_1\alpha_2=0$ and Eq.~\ref{e:homon} in the
case of positive $\alpha_1$ and $\alpha_2$, so assume that
$\alpha_1\alpha_2\neq 0$. Let us rewrite
Eq.~\ref{e:chdergibbs} in the form \be\nonumber \dfrac
{dH_{G}(P^tf_0 )}{dt}= \dfrac{\gamma}{\det Q(t)}e^{-2\gamma
t}h_2(t),\label{e:hcgibbs}\ee where now
$$
h_2(t)=-\alpha_1\alpha_2-\sigma_*e^{-2\lambda
t}\left(\alpha_1\lambda^4 t^2 +\alpha_2(1+\lambda t)^2\right).
$$
Then
$$
h_2'(t)=2\lambda^2\sigma_*e^{-2\lambda
t}t\left(\lambda(\lambda^2\alpha_1+\alpha_2) t
-(\lambda^2\alpha_1-\alpha_2)\right).
$$
Note that for $$ t_*= \dfrac{\lambda^2\alpha_1-\alpha_2}{\lambda
(\lambda^2\alpha_1+\alpha_2)}$$ we have
$$
h_2(t_*)=-\alpha_1(\alpha_2+\lambda^2\sigma_* e^{-2\lambda
t_*}).
$$
A similar analysis as in the overdamped case leads to the
same conclusions so that Eq.~\ref{e:homon1} remains valid
in the case of negative $\alpha_1$ and $\alpha_2$ and also
Eqs.~\ref{e:hopn}-\ref{e:honpn} hold in the same ranges of
parameters in the case of $\alpha_1\alpha_2<0$.

(iii) Finally, let us consider the {\bf underdamped} case
$$
\gamma^2<4\omega^2,
$$
so that  $\lambda_1,\lambda_2$ are complex, and set
$$
\lambda=-\dfrac{\gamma}{2}\qquad \mbox{and}\qquad \beta=\sqrt{
\omega^2-\lambda^2}.
$$
Then $\lambda_1=\lambda+i \beta$ and $\lambda_2=\lambda -i \beta$.
The fundamental matrix in this case is equal to
$$
e^{tF}=\dfrac{e^{\lambda t}}{\beta}\left(%
\begin{array}{cc}
  \beta\cos(\beta t)-\lambda\sin(\beta t) & \sin(\beta t) \\
  -\omega^2\sin(\beta t) & \beta\cos(\beta t)+\lambda\sin(\beta t), \\
\end{array}%
\right).
$$
Let us rewrite the matrix $e^{tF}$ as
$$
e^{tF}=\dfrac{e^{\lambda t}}{\beta}\left(%
\begin{array}{cc}
  c_3(t) & \sin(\beta t) \\
  -\omega^2\sin(\beta t) & c_4(t) \\
\end{array}%
\right),
$$
where
\begin{subequations}\label{e:uhc}
\begin{align}
  c_3(t) &= \beta\cos(\beta t)-\lambda\sin(\beta t), \\
  c_4(t) &= \beta\cos(\beta t)+\lambda\sin(\beta t).
  \end{align}
\end{subequations}
Observe that $\sigma_*$ as defined in Eq.~\ref{e:nsigma} is
equal to $-\sigma^2/4\lambda\omega^2$. The covariance
matrix $R(t)$ is equal to
\begin{equation}\label{e:cmatrixRt}\nonumber
R_*
-\dfrac{\sigma_*e^{2\lambda t}}{\beta^2}\left(%
\begin{array}{cc}
 c_3^2(t)+\omega^2\sin^2(\beta t) &  2\lambda\omega^2\sin^2(\beta t) \\
  2\lambda\omega^2\sin^2(\beta t) & \omega^4\sin^2(\beta t)+\omega^2c_4^2(t) \\
\end{array}%
\right).
\end{equation}
Further
\begin{widetext}
\begin{equation}\label{e:fmatrixQ0}\nonumber
e^{tF}\left(%
\begin{array}{cc}
  \bar\sigma_1^2 & 0 \\
 0 & \bar\sigma_2^2 \\
\end{array}%
\right)e^{tF^T}=
\dfrac{e^{2\lambda t}}{\beta^2}\left(%
\begin{array}{cc}
 \bar\sigma_1^2c_3^2(t)+\bar\sigma_2^2\sin^2(\beta t)
&  \left(-\omega^2\bar\sigma_1^2c_3(t)+\bar\sigma_2^2c_4(t)\right)\sin(\beta t) \\
  \left(-\omega^2\bar\sigma_1^2c_3(t)+\bar\sigma_2^2c_4(t)\right)\sin(\beta t)
& \omega^4\bar\sigma_1^2\sin^2(\beta t)+\bar\sigma_2^2c_4^2(t) \\
\end{array}%
\right).
\end{equation}
\end{widetext}
Making use of expressions~\ref{e:nalpha} and~\ref{e:uhc},
the sum of the matrices in the two preceding equations
gives
\begin{widetext}
$$
Q(t)=\left(%
\begin{array}{cc}
\sigma_*+ \dfrac{e^{2\lambda
t}}{\beta^2}\left(\alpha_1c_3^2(t)+\alpha_2\sin^2(\beta t)\right) &
\dfrac{e^{2\lambda
t}}{\beta^2}\sin(\beta t)\left(\alpha_2c_4(t)-\omega^2\alpha_1c_3(t)\right)\\
\dfrac{e^{2\lambda t}}{\beta^2} \sin(\beta
t)\left(\alpha_2c_4(t)-\omega^2\alpha_1c_3(t)\right) &
\sigma_*\omega^2+\dfrac{e^{2\lambda
t}}{\beta^2}\left(\omega^4\alpha_1\sin^2(\beta t)+\alpha_2c_4^2(t)\right)\\
\end{array}%
\right),
$$
\end{widetext}
which after some algebra leads to
\begin{align}
\det Q(t)&=\omega^2\sigma_*^2+e^{4\lambda
t}\alpha_1\alpha_2 +\dfrac{\sigma_*e^{2\lambda
t}}{\beta^2}\bigl(\omega^2(\omega^2\alpha_1+\alpha_2)\nonumber \\
\label{e:cvdetQsimp}
&\quad -\lambda^2(\omega^2\alpha_1+\alpha_2)\cos(2\beta t)\\
&\quad -\lambda\beta(\omega^2\alpha_1-\alpha_2)\sin(2\beta
t)\bigr).\nonumber
\end{align}
We have
\begin{equation}\label{e:uhdergibbs}
\begin{split}\
\dfrac {dH_{G}(P^tf_0 )}{dt}&=\dfrac{2\lambda}{\det Q(t)} e^{2\lambda
t}\bigl(\alpha_1\alpha_2e^{2\lambda t}\\
&\quad+\dfrac{\sigma_*}{\beta^2}\left(\omega^4\alpha_1\sin^2(\beta
t)+\alpha_2c_4^2(t)\right)\bigr).
\end{split}
\end{equation}
Since $\lambda=-\gamma/2$, Eq.~\ref{e:ohdergibbs0} holds.
Again, Eq.~\ref{e:uhdergibbs} implies Eq.~\ref{e:hozero} in
the case of $\alpha_1\alpha_2=0$. In the case of positive
$\alpha_1$ and $\alpha_2$ the Gibbs entropy is decreasing.
This corresponds to
\begin{equation}\label{e:ohpard}
  \bar\sigma_1^2>\sigma_*\quad\mbox{and}\quad
  \bar\sigma_2^2>\omega^2\sigma_*,
\end{equation}
and is illustrated in Fig.~\ref{f:ohmon4a}.
\begin{figure}[ht]
\includegraphics[scale=0.5]{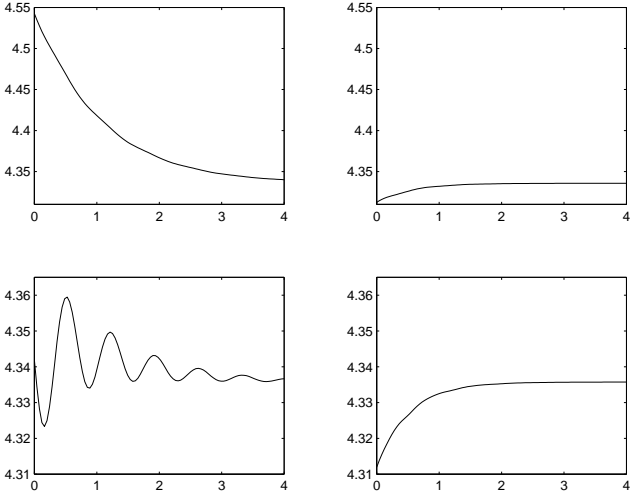}
\caption{Underdamped noisy harmonic oscillator. Plots as in
Fig.~\ref{f:ohmon} but for parameters $\gamma=1$,
$\omega^2={20}$, $\sigma_*=1$. Upper panels are for the
range of parameters as in Eq.~\ref{e:ohpard} with
$\bar\sigma_1=1.1$ and $\bar\sigma_2=5$, lower panels for
Eq.~\ref{e:uhpar1} with $\bar\sigma_1=0.9$,
$\bar\sigma_2=5$.}\label{f:ohmon4a}
\end{figure}

Let us rewrite Eq.~\ref{e:uhdergibbs} in the form
\be\nonumber \dfrac {dH_{G}(P^tf_0 )}{dt}=
\dfrac{\gamma}{\det Q(t)}e^{4 \lambda
t}h_3(t),\label{e:uhcgibbs}\ee where now
$$
h_3(t)=-\alpha_1\alpha_2-\dfrac{\sigma_*}{\beta^2}e^{-2\lambda
t}\left(\omega^4\alpha_1\sin^2(\beta t)+\alpha_2c_4^2(t)\right).
$$
We have
\begin{equation}\nonumber
\begin{split}
h_3'(t)&=\dfrac{2\omega^2\sigma_*}{\beta^2}e^{-2\lambda t}\sin(\beta
t)\bigl(\lambda(\omega^2\alpha_1 +\alpha_2)\sin(\beta t)
\\
&\quad+\beta(\alpha_2-\omega^2\alpha_1)\cos(\beta t)\bigr).
\end{split}
\end{equation}
The function $h_3$ has extreme values at all $t$ for which either
$\sin(\beta t)=0$ or
\begin{equation}\label{e:uhzeros}
\beta\cos(\beta t)=\lambda\sin(\beta
t)\dfrac{(\omega^2\alpha_1+\alpha_2)}{(\omega^2\alpha_1-\alpha_2)}.
\end{equation}
Making use of the relations $\omega^2=\lambda^2+\beta^2$
and $\lambda=-\gamma/2$ it is seen that for every
nonnegative integer $k$ we have
\begin{equation}\label{e:uhexval1}
h_3(k\pi/\beta)=-\alpha_2(\alpha_1+\sigma_*e^{\gamma
k\pi/\beta})
\end{equation}
and
\begin{equation}\label{e:uhexval2}
h_3(t_*+k\pi/\beta)=
-\alpha_1\left(\alpha_2+\sigma_*\omega^2 e^{\gamma (t_*
+k\pi/\beta)}\right),
\end{equation}
where $t_*$ is the smallest positive solution of
Eq.~\ref{e:uhzeros}.  Thus, if $\alpha_1<0$ and
$\alpha_2>0$ then $h_3(k\pi/\beta)<0$ and
$h_3(t_*+k\pi/\beta)>0$ for all $k$. Consequently, if
\begin{equation}\label{e:uhpar1}
  \bar\sigma_1^2<\sigma_*\quad\mbox{and}\quad
  \bar\sigma_2^2>\omega^2\sigma_*,
\end{equation}
then there are two infinite sequences of points $t_k$ and $\bar t_k$
such that \be \dfrac {dH_{G}(P^tf_0 )}{dt}\left \{
    \begin{array}{ll}
<& 0   \quad \mbox{for} \quad t_{k}<t< \bar t_k, \\
>& 0   \quad \mbox{for} \quad \bar t_k <t< t_{k+1}.
\end{array} \right.\label{e:uhosc} \ee

Consider now the case of $\alpha_2<0$. From
Eq.~\ref{e:uhexval1} it follows that $h_3(k\pi/\beta)>0$.
When $\alpha_1<0$ the values of $h_3$ at $t_*+k\pi/\beta$
are positive. Therefore $h_3(t)>0$ for all $t>0$.
Consequently, if
\begin{equation}\label{e:uhpar2}
  \bar\sigma_1^2<\sigma_*\quad\mbox{and}\quad
  \bar\sigma_2^2<\omega^2\sigma_*,
\end{equation}
then the Gibbs' entropy increases. Finally, when
$\alpha_1>0$ then $\alpha_2<\omega^2\alpha_1$, the function
$h_3$ decreases from a positive value at $k\pi/\beta$ to a
negative value at $t_*+k\pi/\beta$ and then increases back
to a positive value. Consequently, if
\begin{equation}\label{e:uhpar3}
  \bar\sigma_1^2>\sigma_*\quad\mbox{and}\quad
  \bar\sigma_2^2<\omega^2\sigma_*,
\end{equation}
then there are two infinite sequences of points $t_k$ and
$\bar t_k$ such that
    \be \dfrac {dH_{G}(P^tf_0 )}{dt}\left\{
    \begin{array}{ll}
>& 0   \quad \mbox{for} \quad t_{k}<t< \bar t_k, \\
<& 0   \quad \mbox{for} \quad \bar t_k <t< t_{k+1}.
\end{array} \right.\label{e:uhosc1} \ee
These behaviors are illustrated in Fig.~\ref{f:ohmon5}.
\begin{figure}[ht]
\includegraphics[scale=0.5]{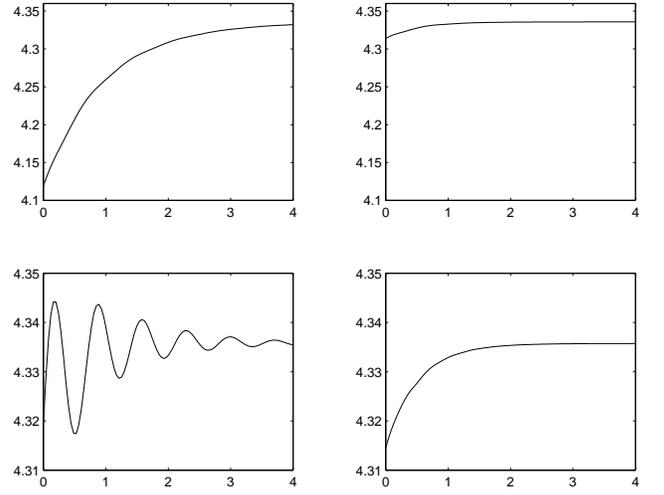}
\caption{Underdamped noisy harmonic oscillator. Plots as in
Fig.~\ref{f:ohmon4a}. Upper panels are for the range of
parameters as in Eq.~\ref{e:uhpar2} with $\bar\sigma_1=0.9$
and $\bar\sigma_2=4$, lower panels for Eq.~\ref{e:uhpar3}
with $\bar\sigma_1=1.1$, $\bar\sigma_2=4$.
}\label{f:ohmon5}\end{figure}

\end{example}

\setcounter{equation}{0}
\section{Summary and Discussion}\label{s:disc}

From the most general properties of the conditional
entropy, it may remain constant or increase
(Theorem~\ref{t:voigt}). In invertible systems (e.g.
measure preserving systems of differential equations or
invertible maps) the conditional entropy is fixed at the
value with which the system is prepared
(Theorem~\ref{thm-invert}, see also
\citep{andrey85,steeb79,mcmtdbk,mcmitaly}).  This property
is  illustrated by Example~\ref{exp:rotation}. The addition
of noise can reverse this invertibility property and induce
the dynamic property of asymptotic stability. Asymptotic
stability  is necessary and sufficient for the monotonic
evolution of the conditional entropy to a maximum value of
zero, c.f. Theorem~\ref{t:entropyconv}. This has been amply
and fully illustrated in Examples~\ref{ex-o-u}
and~\ref{sss:ho} for the Ornstein-Uhlenbeck process and a
noisy harmonic oscillator respectively.

The situation is much different for the Gibbs' entropy,
however, and it is often difficult to make general
statements about what the temporal behavior will be
\citep{steeb79,andrey85}.  The  rate of change of the
Gibbs' entropy in invertible systems depends on the
Jacobian (c.f. Eq.~\ref{gibbsrate}), and it is only for
Lebesque measure preserving flows that the Gibbs' entropy
is constant.   In considering the Gibbs' entropy in
invertible systems,  Example~\ref{exp:ct-friction} treats a
general two dimensional system.  When  the steady state is
stable $(\lambda_1 + \lambda_2 < 0)$ then the Gibbs'
entropy diverges to  $-\infty$. Alternately, when the
steady state is unstable  $(\lambda_1 + \lambda_2 > 0)$
then the Gibbs' entropy diverges to  $\infty$.
Example~\ref{exp:ct-ho} considers the specific case of a
damped harmonic oscillator in which the Gibbs entropy
diverges to $-\infty$.  Example~\ref{exp:rotation} treats
the measure preserving rotation on the circle and shows
that the Gibbs' entropy is constant and fixed at the value
corresponding to the way in which the system was
prepared--as is the conditional entropy.

The situation with the temporal behavior of the Gibbs'
entropy becomes even more curious when an invertible system
is subjected to noise and thus rendered non-invertible.  A
number of authors have considered aspects of this recently,
notably \citet{ruelle96,ruelle03}, \citet{nicolis98},
\citet{nicolis99}, 
\citet{bag00,bag01,bag02,bag02b,bag03}, and \citet{garb05}. As we have shown in
Example~\ref{ex-o-u}, in contrast to the conditional
entropy that increases monotonically to approach zero, the
Gibbs' entropy monotonically approaches the equilibrium
value of $H_G(f_*)$ by either increasing or decreasing and
the direction of movement is totally determined by the
variance $ \bar \sigma ^2$ of the initial ensemble.  The
temporal behavior of the Gibbs' entropy can, however, have
even more complicated patterns as illustrated by
Example~\ref{sss:ho}.  There, we have
shown that when the harmonic oscillator is either over
damped or critically damped that the approach of
$H_G(P^tf_0 )$ to $H_G(f_*)$ may be either  monotonic
increasing or increasing (Fig.~\ref{f:ohmon}), or approach
the equilibrium value with an undershoot or overshoot
(Figs.~\ref{f:ohmon1} and~\ref{f:ohmon2}).  When the
harmonic oscillator is under damped then the approach of
the Gibbs' entropy to $H_G(f_*)$ may even be oscillatory as
shown in Figs.~\ref{f:ohmon4a} and~\ref{f:ohmon5}. All of
these patterns of temporal behavior are, as we have shown,
totally dependent on the relation of the variance of the
initial ensemble to the variance of the equilibrium state.
Remember that in all of these cases (over, critically, and
under damped) the conditional entropy smoothly approaches
zero so $H_c(P^tf_0 |f_*) + H_G(f_*)$, as shown in the
right hand panels of Figs.~\ref{f:ohmon}
through~\ref{f:ohmon5}, monotonically increases to approach
$H_G(f_*)$.

The concept of entropy originally arose in the context of
the second law of thermodynamics.   Following
\citet{landau80}, we may formulate
 the second law of thermodynamics as follows.
 Let $S_{TD}(t) $ be
defined as the time dependent thermodynamic entropy. Then
for an isolated system
\begin{equation}
S_{TD}(t_2) \geq S_{TD}(t_1) \qquad \mbox{for all} \quad
t_2>t_1, \label{e-second}
\end{equation}
and there is a unique steady state
\begin{equation}
S_{TD}^*   = \lim_{t\to +\infty} S_{TD}(t) \label{e-entss}
\end{equation}
for all initial system preparations.  The entropy
difference satisfies
\begin{equation}
\Delta S(t) \equiv S_{TD}(t) - S_{TD}^* \leq 0
\label{e-entdiff}
\end{equation}
and
\begin{equation}
\lim_{t\to +\infty} \Delta S(t) = 0.
\end{equation}
In other words, the system entropy evolves  to a unique
maximum for all system preparations.

In attempts to give a dynamical interpretation of the
second  law, it is assumed that a thermodynamic  system has
states distributed in the phase  space $ {X}$. The
distribution of these states is characterized by a (time
dependent) density ${f(t,x)}$. A thermodynamic equilibrium
is assumed be characterized by a stationary (time
independent) density $ {f_*(x)}$.

The Gibbs' equilibrium entropy definition Eq.~\ref{e-gibbs}
has repeatedly proven to yield correct results when applied
to a variety of equilibrium situations.  This is why it is
the gold standard for equilibrium computations in
statistical mechanics and thermodynamics. Thus it makes
total sense to  identify the equilibrium  Gibbs' entropy
${H_G(f_*)}$ with the equilibrium  thermodynamic entropy $
{S_{TD}^*}$
$$
S_{TD}^* \equiv   H_G(f_*).
$$

Do the results  in
Sections~\ref{s:det} and~\ref{s:noninvert} on the dynamic
behavior of  the conditional and Gibbs' entropies that we have
determined  analytically, and  illustrated with examples,
offer any insight into  dynamic analogs of $S_{TD}(t)$ and
$\Delta S(t)$?

The question of how a time dependent non-equilibrium
entropy should be defined has interested  investigators for
some time, and specifically the question of whether the
Gibbs' entropy $H_G(f)$ can be taken to coincide with the
time dependent entropy $S_{TD}(t)$ of the second law of
thermodynamics has occupied many researchers.  Various
aspects of this question have been considered in
\citep{nicolis99,bag00,bag01,bag02,bag02b,bag03,bag04,ruelle96,ruelle97,ruelle03,ruelle04}.

The  non-equilibrium Gibbs' entropy $H_G(f)$ is manifestly
not a  good candidate for $S_{TD}(t)$ because its dynamical
behavior is at odds with what is demanded by the Second Law
of Thermodynamics.  As we have demonstrated and summarized,
concrete examples can be constructed in which the direction
of the temporal change in $H_G(f)$ depends on the initial
preparation of the system and others can be constructed in
which $H_G(f)$ oscillates in time.  Thus there is good
reason to search for a different analog of $S_{TD}(t)$.

A number of authors, among them \citet[pp. 122-129, Eq.
247]{degroot84},  \citet[pp. 111-114 and 185]{vankampen92},
and \citet[p. 213]{penrose05}  have suggested that
 $S_{TD}(t)$ should be associated dynamically with
      \bea
    H_{NE}(f) &\equiv& H_c(f|f_*e^{H_G(f_*)}) \nonumber \\
    &=& H_c(f|f_*) + H_G(f_*)
    \eea
as an extension of  \citet[pp. 44-45 and 168]{gibbs02}
discussion  of entropy.  This also goes under the name of
the ``Gibbs' entropy postulate"
\citep{mazur94,mazur98a,mazur98b,mazur00,mazur01,rubi01}.

Here, we have shown that $H_{NE}(f) = H_c(f|f_*) +
H_G(f_*)$ has  the temporal behavior required for the
entropy   $S_{TD}(t)$ of the second law of thermodynamics.
This is a consequence of the temporal behavior of
$H_c(f|f_*)$.  Namely $H_{NE}(f)$ is either constant for
invertible dynamics,  or monotone increasing to the
equilibrium value of  $H_G(f_*)$ for non-invertible
asymptotically stable dynamics induced by noise
perturbations.  Once this identification is granted, then
it follows that  $\Delta S(t)$ should be identified  with
$H_c(f|f_*)$: \be \Delta S(t) \equiv H_c(f|f_*)=-\int_X
f(t,x)\log\dfrac{f(t,x)}{f_*(x)}dx\label{deltas}, \ee as
has been previously suggested \citep{mcm89rmp,mcmtdbk}.

\begin{acknowledgments}
This work was supported by the Natural Sciences and Engineering Research Council (NSERC grant OGP-0036920,
Canada) and MITACS. This research was carried out while MT-K was visiting McGill University in 2004 and 2005.
\end{acknowledgments}

\bibliography{zpf}

\begin{thebibliography}{58}
\expandafter\ifx\csname natexlab\endcsname\relax\def\natexlab#1{#1}\fi
\expandafter\ifx\csname bibnamefont\endcsname\relax
  \def\bibnamefont#1{#1}\fi
\expandafter\ifx\csname bibfnamefont\endcsname\relax
  \def\bibfnamefont#1{#1}\fi
\expandafter\ifx\csname citenamefont\endcsname\relax
  \def\citenamefont#1{#1}\fi
\expandafter\ifx\csname url\endcsname\relax
  \def\url#1{\texttt{#1}}\fi
\expandafter\ifx\csname urlprefix\endcsname\relax\def\urlprefix{URL }\fi
\providecommand{\bibinfo}[2]{#2}
\providecommand{\eprint}[2][]{\url{#2}}

\bibitem[{\citenamefont{Loskot and Rudnicki}(1991)}]{loskot91}
\bibinfo{author}{\bibfnamefont{K.}~\bibnamefont{Loskot}} \bibnamefont{and}
  \bibinfo{author}{\bibfnamefont{R.}~\bibnamefont{Rudnicki}},
  \bibinfo{journal}{Ann. Pol. Math.} \textbf{\bibinfo{volume}{52}},
  \bibinfo{pages}{140} (\bibinfo{year}{1991}).

\bibitem[{\citenamefont{Abbondandolo}(1999)}]{abbond99}
\bibinfo{author}{\bibfnamefont{A.}~\bibnamefont{Abbondandolo}},
  \bibinfo{journal}{Stoch. Anal. Applic.} \textbf{\bibinfo{volume}{17}},
  \bibinfo{pages}{131} (\bibinfo{year}{1999}).

\bibitem[{\citenamefont{Toscani and Villani}(2000)}]{toscanivillani}
\bibinfo{author}{\bibfnamefont{G.}~\bibnamefont{Toscani}} \bibnamefont{and}
  \bibinfo{author}{\bibfnamefont{C.}~\bibnamefont{Villani}},
  \bibinfo{journal}{J. Stat. Phys.} \textbf{\bibinfo{volume}{98}},
  \bibinfo{pages}{1279} (\bibinfo{year}{2000}).

\bibitem[{\citenamefont{Arnold et~al.}(2001)\citenamefont{Arnold, Markowich,
  Toscani, and Unterreiter}}]{arnold01}
\bibinfo{author}{\bibfnamefont{A.}~\bibnamefont{Arnold}},
  \bibinfo{author}{\bibfnamefont{P.}~\bibnamefont{Markowich}},
  \bibinfo{author}{\bibfnamefont{G.}~\bibnamefont{Toscani}}, \bibnamefont{and}
  \bibinfo{author}{\bibfnamefont{A.}~\bibnamefont{Unterreiter}},
  \bibinfo{journal}{Comm. Partial Differential Equations}
  \textbf{\bibinfo{volume}{26}}, \bibinfo{pages}{43} (\bibinfo{year}{2001}).

\bibitem[{\citenamefont{Qian}(2001)}]{qian01}
\bibinfo{author}{\bibfnamefont{H.}~\bibnamefont{Qian}}, \bibinfo{journal}{Phys.
  Rev. E.} \textbf{\bibinfo{volume}{63}}, \bibinfo{pages}{042103}
  (\bibinfo{year}{2001}).

\bibitem[{\citenamefont{Qian et~al.}(2002)\citenamefont{Qian, Qian, and
  Tang}}]{qian02}
\bibinfo{author}{\bibfnamefont{H.}~\bibnamefont{Qian}},
  \bibinfo{author}{\bibfnamefont{M.}~\bibnamefont{Qian}}, \bibnamefont{and}
  \bibinfo{author}{\bibfnamefont{X.}~\bibnamefont{Tang}}, \bibinfo{journal}{J.
  Stat. Phys.} \textbf{\bibinfo{volume}{107}}, \bibinfo{pages}{1129}
  (\bibinfo{year}{2002}).

\bibitem[{\citenamefont{Markowich and Villani}(2000)}]{markowich00}
\bibinfo{author}{\bibfnamefont{P.}~\bibnamefont{Markowich}} \bibnamefont{and}
  \bibinfo{author}{\bibfnamefont{C.}~\bibnamefont{Villani}},
  \bibinfo{journal}{Mat. Contemp.} \textbf{\bibinfo{volume}{19}},
  \bibinfo{pages}{1} (\bibinfo{year}{2000}).

\bibitem[{\citenamefont{Gibbs}(1962)}]{gibbs02}
\bibinfo{author}{\bibfnamefont{J.}~\bibnamefont{Gibbs}},
  \emph{\bibinfo{title}{Elementary Principles in Statistical Mechanics}}
  (\bibinfo{publisher}{Dover}, \bibinfo{address}{New York},
  \bibinfo{year}{1962}).

\bibitem[{\citenamefont{Ruelle}(1996)}]{ruelle96}
\bibinfo{author}{\bibfnamefont{D.}~\bibnamefont{Ruelle}}, \bibinfo{journal}{J.
  Stat. Phys.} \textbf{\bibinfo{volume}{85}}, \bibinfo{pages}{1}
  (\bibinfo{year}{1996}).

\bibitem[{\citenamefont{Ruelle}(2003)}]{ruelle03}
\bibinfo{author}{\bibfnamefont{D.}~\bibnamefont{Ruelle}},
  \bibinfo{journal}{Proc. Nat. Acad. Sci.} \textbf{\bibinfo{volume}{100}},
  \bibinfo{pages}{3054} (\bibinfo{year}{2003}).

\bibitem[{\citenamefont{Nicolis and Daems}(1998)}]{nicolis98}
\bibinfo{author}{\bibfnamefont{G.}~\bibnamefont{Nicolis}} \bibnamefont{and}
  \bibinfo{author}{\bibfnamefont{D.}~\bibnamefont{Daems}},
  \bibinfo{journal}{Chaos} \textbf{\bibinfo{volume}{8}}, \bibinfo{pages}{311}
  (\bibinfo{year}{1998}).

\bibitem[{\citenamefont{Daems and Nicolis}(1999)}]{nicolis99}
\bibinfo{author}{\bibfnamefont{D.}~\bibnamefont{Daems}} \bibnamefont{and}
  \bibinfo{author}{\bibfnamefont{G.}~\bibnamefont{Nicolis}},
  \bibinfo{journal}{Phys. Rev. E.} \textbf{\bibinfo{volume}{59}},
  \bibinfo{pages}{4000} (\bibinfo{year}{1999}).

\bibitem[{\citenamefont{Bag et~al.}(2000)\citenamefont{Bag, Chaudhuri, and
  Ray}}]{bag00}
\bibinfo{author}{\bibfnamefont{B.}~\bibnamefont{Bag}},
  \bibinfo{author}{\bibfnamefont{J.}~\bibnamefont{Chaudhuri}},
  \bibnamefont{and} \bibinfo{author}{\bibfnamefont{D.}~\bibnamefont{Ray}},
  \bibinfo{journal}{J. Phys. A} \textbf{\bibinfo{volume}{33}},
  \bibinfo{pages}{8331} (\bibinfo{year}{2000}).

\bibitem[{\citenamefont{Bag et~al.}(2001)\citenamefont{Bag, Banik, and
  Ray}}]{bag01}
\bibinfo{author}{\bibfnamefont{B.}~\bibnamefont{Bag}},
  \bibinfo{author}{\bibfnamefont{S.}~\bibnamefont{Banik}}, \bibnamefont{and}
  \bibinfo{author}{\bibfnamefont{D.}~\bibnamefont{Ray}},
  \bibinfo{journal}{Phys. Rev. E.} \textbf{\bibinfo{volume}{64}},
  \bibinfo{pages}{026110} (\bibinfo{year}{2001}).

\bibitem[{\citenamefont{Bag}(2002{\natexlab{a}})}]{bag02}
\bibinfo{author}{\bibfnamefont{B.}~\bibnamefont{Bag}}, \bibinfo{journal}{Phys.
  Rev. E} \textbf{\bibinfo{volume}{66}}, \bibinfo{pages}{026122}
  (\bibinfo{year}{2002}{\natexlab{a}}).

\bibitem[{\citenamefont{Bag}(2002{\natexlab{b}})}]{bag02b}
\bibinfo{author}{\bibfnamefont{B.}~\bibnamefont{Bag}}, \bibinfo{journal}{Phys.
  Rev. E} \textbf{\bibinfo{volume}{65}}, \bibinfo{pages}{046118}
  (\bibinfo{year}{2002}{\natexlab{b}}).

\bibitem[{\citenamefont{Bag}(2003)}]{bag03}
\bibinfo{author}{\bibfnamefont{B.}~\bibnamefont{Bag}}, \bibinfo{journal}{J.
  Chem. Phys.} \textbf{\bibinfo{volume}{119}}, \bibinfo{pages}{4988}
  (\bibinfo{year}{2003}).

\bibitem[{\citenamefont{Mackey and Tyran-Kami\'nska}(2005)}]{mackeytyran05}
\bibinfo{author}{\bibfnamefont{M.~C.} \bibnamefont{Mackey}} \bibnamefont{and}
  \bibinfo{author}{\bibfnamefont{M.}~\bibnamefont{Tyran-Kami\'nska}},
  \bibinfo{journal}{cond-mat/0501092}  (\bibinfo{year}{2005}).

\bibitem[{\citenamefont{Khinchin}(1949)}]{khinchin49}
\bibinfo{author}{\bibfnamefont{A.}~\bibnamefont{Khinchin}},
  \emph{\bibinfo{title}{Mathematical Foundations of Statistical Mechanics}}
  (\bibinfo{publisher}{Dover}, \bibinfo{address}{New York},
  \bibinfo{year}{1949}).

\bibitem[{\citenamefont{Skagerstam}(1974)}]{skagerstam74}
\bibinfo{author}{\bibfnamefont{B.}~\bibnamefont{Skagerstam}},
  \bibinfo{journal}{Z. Naturforsch.} \textbf{\bibinfo{volume}{29A}},
  \bibinfo{pages}{1239} (\bibinfo{year}{1974}).

\bibitem[{\citenamefont{Lasota and Mackey}(1994)}]{almcmbk94}
\bibinfo{author}{\bibfnamefont{A.}~\bibnamefont{Lasota}} \bibnamefont{and}
  \bibinfo{author}{\bibfnamefont{M.~C.} \bibnamefont{Mackey}},
  \emph{\bibinfo{title}{Chaos, Fractals and Noise: Stochastic Aspects of
  Dynamics}} (\bibinfo{publisher}{Springer-Verlag}, \bibinfo{address}{Berlin,
  New York, Heidelberg}, \bibinfo{year}{1994}).

\bibitem[{\citenamefont{Eu}(1995)}]{eu95}
\bibinfo{author}{\bibfnamefont{B.~C.} \bibnamefont{Eu}}, \bibinfo{journal}{J.
  Chem. Phys.} \textbf{\bibinfo{volume}{102}} (\bibinfo{year}{1995}).

\bibitem[{\citenamefont{Eu}(1997)}]{eu97}
\bibinfo{author}{\bibfnamefont{B.~C.} \bibnamefont{Eu}}, \bibinfo{journal}{J.
  Chem. Phys.} \textbf{\bibinfo{volume}{106}} (\bibinfo{year}{1997}).

\bibitem[{\citenamefont{D-q Jian and p~Qian}(2000)}]{jiang00}
\bibinfo{author}{\bibfnamefont{M.~Q.} \bibnamefont{D-q Jian}} \bibnamefont{and}
  \bibinfo{author}{\bibfnamefont{M.}~\bibnamefont{p~Qian}},
  \bibinfo{journal}{Commun. Math. Phys.} \textbf{\bibinfo{volume}{214}}
  (\bibinfo{year}{2000}).

\bibitem[{\citenamefont{Qian}(2002)}]{qian02a}
\bibinfo{author}{\bibfnamefont{H.}~\bibnamefont{Qian}}, \bibinfo{journal}{J.
  Phys. Chem.} \textbf{\bibinfo{volume}{106}}, \bibinfo{pages}{2065}
  (\bibinfo{year}{2002}).

\bibitem[{\citenamefont{Sachs}(1987)}]{sachs87}
\bibinfo{author}{\bibfnamefont{R.}~\bibnamefont{Sachs}},
  \emph{\bibinfo{title}{The Physics of Time Reversal}}
  (\bibinfo{publisher}{University of Chicago Press},
  \bibinfo{address}{Chicago}, \bibinfo{year}{1987}).

\bibitem[{\citenamefont{Reichenbach}(1957)}]{reichenbach57}
\bibinfo{author}{\bibfnamefont{H.}~\bibnamefont{Reichenbach}},
  \emph{\bibinfo{title}{The Direction of Time}} (\bibinfo{publisher}{California
  University Press}, \bibinfo{address}{Berekeley}, \bibinfo{year}{1957}).

\bibitem[{\citenamefont{Steeb}(1979)}]{steeb79}
\bibinfo{author}{\bibfnamefont{W.-H.} \bibnamefont{Steeb}},
  \bibinfo{journal}{Physica A} \textbf{\bibinfo{volume}{95}},
  \bibinfo{pages}{181} (\bibinfo{year}{1979}).

\bibitem[{\citenamefont{Andrey}(1985)}]{andrey85}
\bibinfo{author}{\bibfnamefont{L.}~\bibnamefont{Andrey}},
  \bibinfo{journal}{Physics Letters A} \textbf{\bibinfo{volume}{111}},
  \bibinfo{pages}{45} (\bibinfo{year}{1985}).

\bibitem[{\citenamefont{Voigt}(1981)}]{voigt81}
\bibinfo{author}{\bibfnamefont{J.}~\bibnamefont{Voigt}},
  \bibinfo{journal}{Commun. Math. Phys.} \textbf{\bibinfo{volume}{81}},
  \bibinfo{pages}{31} (\bibinfo{year}{1981}).

\bibitem[{\citenamefont{Csisz{\'a}r}(1975)}]{csiszar75}
\bibinfo{author}{\bibfnamefont{I.}~\bibnamefont{Csisz{\'a}r}},
  \bibinfo{journal}{Ann. Probability} \textbf{\bibinfo{volume}{3}},
  \bibinfo{pages}{146} (\bibinfo{year}{1975}).

\bibitem[{\citenamefont{Ruelle}(1976)}]{ruelle76}
\bibinfo{author}{\bibfnamefont{D.}~\bibnamefont{Ruelle}}, \bibinfo{journal}{Am.
  J. Math.} \textbf{\bibinfo{volume}{98}}, \bibinfo{pages}{619}
  (\bibinfo{year}{1976}).

\bibitem[{\citenamefont{Ruelle}(1980)}]{ruelle80}
\bibinfo{author}{\bibfnamefont{D.}~\bibnamefont{Ruelle}},
  \bibinfo{journal}{Ann. N.Y. Acad. Sci.} \textbf{\bibinfo{volume}{357}},
  \bibinfo{pages}{1} (\bibinfo{year}{1980}).

\bibitem[{\citenamefont{D.J.~Evans and Morris}(1990)}]{evans90}
\bibinfo{author}{\bibfnamefont{E.~C.} \bibnamefont{D.J.~Evans}}
  \bibnamefont{and} \bibinfo{author}{\bibfnamefont{G.}~\bibnamefont{Morris}},
  \bibinfo{journal}{Phys. Rev. A} \textbf{\bibinfo{volume}{42}},
  \bibinfo{pages}{5990} (\bibinfo{year}{1990}).

\bibitem[{\citenamefont{D.J.~Evans and Morris}(1993)}]{evans93}
\bibinfo{author}{\bibfnamefont{E.~C.} \bibnamefont{D.J.~Evans}}
  \bibnamefont{and} \bibinfo{author}{\bibfnamefont{G.}~\bibnamefont{Morris}},
  \bibinfo{journal}{Phys. Rev. Let.} \textbf{\bibinfo{volume}{71}},
  \bibinfo{pages}{2401} (\bibinfo{year}{1993}).

\bibitem[{\citenamefont{Gallavotti}(1999)}]{gallavotti}
\bibinfo{author}{\bibfnamefont{G.}~\bibnamefont{Gallavotti}},
  \emph{\bibinfo{title}{Statistical Mechanics: A Short Treatise}}
  (\bibinfo{publisher}{Springer Verlag}, \bibinfo{address}{Berlin, New York},
  \bibinfo{year}{1999}).

\bibitem[{\citenamefont{Horsthemke and Lefever}(1984)}]{horsthemke84}
\bibinfo{author}{\bibfnamefont{W.}~\bibnamefont{Horsthemke}} \bibnamefont{and}
  \bibinfo{author}{\bibfnamefont{R.}~\bibnamefont{Lefever}},
  \emph{\bibinfo{title}{Noise Induced Transitions: Theory and Applications in
  Physics, Chemistry, and Biology}} (\bibinfo{publisher}{Springer-Verlag},
  \bibinfo{address}{Berlin, New York, Heidelberg}, \bibinfo{year}{1984}).

\bibitem[{\citenamefont{Risken}(1984)}]{risken84}
\bibinfo{author}{\bibfnamefont{H.}~\bibnamefont{Risken}},
  \emph{\bibinfo{title}{The Fokker-Planck Equation}}
  (\bibinfo{publisher}{Springer-Verlag}, \bibinfo{address}{Berlin, New York,
  Heidelberg}, \bibinfo{year}{1984}).

\bibitem[{\citenamefont{Majee and Bag}(2004)}]{bag04}
\bibinfo{author}{\bibfnamefont{P.}~\bibnamefont{Majee}} \bibnamefont{and}
  \bibinfo{author}{\bibfnamefont{B.}~\bibnamefont{Bag}}, \bibinfo{journal}{J.
  Phys. A.} \textbf{\bibinfo{volume}{37}}, \bibinfo{pages}{3353}
  (\bibinfo{year}{2004}).

\bibitem[{\citenamefont{Zakai and Snyders}(1970)}]{zakaisnyders70}
\bibinfo{author}{\bibfnamefont{M.}~\bibnamefont{Zakai}} \bibnamefont{and}
  \bibinfo{author}{\bibfnamefont{J.}~\bibnamefont{Snyders}},
  \bibinfo{journal}{J. Differential Equations} \textbf{\bibinfo{volume}{8}},
  \bibinfo{pages}{27} (\bibinfo{year}{1970}).

\bibitem[{\citenamefont{Erickson}(1971)}]{erickson71}
\bibinfo{author}{\bibfnamefont{R.~V.} \bibnamefont{Erickson}},
  \bibinfo{journal}{Ann. Math. Statist.} \textbf{\bibinfo{volume}{42}},
  \bibinfo{pages}{820} (\bibinfo{year}{1971}).

\bibitem[{\citenamefont{Chandrasekhar}(1943)}]{chandra43}
\bibinfo{author}{\bibfnamefont{S.}~\bibnamefont{Chandrasekhar}},
  \bibinfo{journal}{Rev. Mod. Phys.} \textbf{\bibinfo{volume}{15}},
  \bibinfo{pages}{1} (\bibinfo{year}{1943}).

\bibitem[{\citenamefont{Mackey}(1992)}]{mcmtdbk}
\bibinfo{author}{\bibfnamefont{M.~C.} \bibnamefont{Mackey}},
  \emph{\bibinfo{title}{Time's Arrow: The Origins of Thermodynamic Behaviour}}
  (\bibinfo{publisher}{Springer-Verlag}, \bibinfo{address}{Berlin, New York,
  Heidelberg}, \bibinfo{year}{1992}).

\bibitem[{\citenamefont{Mackey}(2001)}]{mcmitaly}
\bibinfo{author}{\bibfnamefont{M.~C.} \bibnamefont{Mackey}}, in
  \emph{\bibinfo{booktitle}{Time's Arrows, Quantum Measurement and Superluminal
  Behavior}}, edited by
  \bibinfo{editor}{\bibfnamefont{C.}~\bibnamefont{Mugnai}},
  \bibinfo{editor}{\bibfnamefont{A.}~\bibnamefont{Ranfagni}}, \bibnamefont{and}
  \bibinfo{editor}{\bibfnamefont{L.}~\bibnamefont{Schulman}}
  (\bibinfo{publisher}{Consiglio Nazionale Delle Richerche},
  \bibinfo{address}{Roma}, \bibinfo{year}{2001}), pp. \bibinfo{pages}{49--65}.

\bibitem[{\citenamefont{Garbaczewski}(2005)}]{garb05}
\bibinfo{author}{\bibfnamefont{P.}~\bibnamefont{Garbaczewski}},
  \bibinfo{journal}{Physics Letters A} \textbf{\bibinfo{volume}{341}},
  \bibinfo{pages}{33} (\bibinfo{year}{2005}).

\bibitem[{\citenamefont{Landau and Lifshitz}(1980)}]{landau80}
\bibinfo{author}{\bibfnamefont{L.}~\bibnamefont{Landau}} \bibnamefont{and}
  \bibinfo{author}{\bibfnamefont{E.}~\bibnamefont{Lifshitz}},
  \emph{\bibinfo{title}{Statistical Physics: Part 1}}
  (\bibinfo{publisher}{Butterworth-Heinemann}, \bibinfo{address}{London},
  \bibinfo{year}{1980}), \bibinfo{edition}{3rd} ed.

\bibitem[{\citenamefont{Ruelle}(1997)}]{ruelle97}
\bibinfo{author}{\bibfnamefont{D.}~\bibnamefont{Ruelle}},
  \bibinfo{journal}{Commun. Math. Phys.} \textbf{\bibinfo{volume}{189}},
  \bibinfo{pages}{365} (\bibinfo{year}{1997}).

\bibitem[{\citenamefont{Ruelle}(2004)}]{ruelle04}
\bibinfo{author}{\bibfnamefont{D.}~\bibnamefont{Ruelle}},
  \bibinfo{journal}{Phys. Today}  (\bibinfo{year}{2004}).

\bibitem[{\citenamefont{de~Groot and Mazur}(1984)}]{degroot84}
\bibinfo{author}{\bibfnamefont{S.}~\bibnamefont{de~Groot}} \bibnamefont{and}
  \bibinfo{author}{\bibfnamefont{P.}~\bibnamefont{Mazur}},
  \emph{\bibinfo{title}{Non-Equilibrium Thermodynamics}}
  (\bibinfo{publisher}{Dover}, \bibinfo{address}{New York},
  \bibinfo{year}{1984}).

\bibitem[{\citenamefont{van Kampen}(1992)}]{vankampen92}
\bibinfo{author}{\bibfnamefont{N.}~\bibnamefont{van Kampen}},
  \emph{\bibinfo{title}{Stochastic Processes in Physics and Chemistry}}
  (\bibinfo{publisher}{Elesvier-North Holland}, \bibinfo{address}{Amsterdam},
  \bibinfo{year}{1992}), \bibinfo{edition}{2nd} ed.

\bibitem[{\citenamefont{Penrose}(2005)}]{penrose05}
\bibinfo{author}{\bibfnamefont{O.}~\bibnamefont{Penrose}},
  \emph{\bibinfo{title}{Foundations of Statistical Mechanics}}
  (\bibinfo{publisher}{Dover}, \bibinfo{address}{Mineola, New York},
  \bibinfo{year}{2005}), \bibinfo{edition}{revised} ed.

\bibitem[{\citenamefont{A.~P{'e}rez-Madrid and Mazur}(1994)}]{mazur94}
\bibinfo{author}{\bibfnamefont{J.~R.} \bibnamefont{A.~P{'e}rez-Madrid}}
  \bibnamefont{and} \bibinfo{author}{\bibfnamefont{P.}~\bibnamefont{Mazur}},
  \bibinfo{journal}{Physica A} \textbf{\bibinfo{volume}{212}},
  \bibinfo{pages}{231} (\bibinfo{year}{1994}).

\bibitem[{\citenamefont{Rub{'i} and Mazur}(1998)}]{mazur98a}
\bibinfo{author}{\bibfnamefont{J.}~\bibnamefont{Rub{'i}}} \bibnamefont{and}
  \bibinfo{author}{\bibfnamefont{P.}~\bibnamefont{Mazur}},
  \bibinfo{journal}{Physica A} \textbf{\bibinfo{volume}{250}},
  \bibinfo{pages}{253} (\bibinfo{year}{1998}).

\bibitem[{\citenamefont{Mazur}(1998)}]{mazur98b}
\bibinfo{author}{\bibfnamefont{P.}~\bibnamefont{Mazur}},
  \bibinfo{journal}{Physica A} \textbf{\bibinfo{volume}{261}},
  \bibinfo{pages}{451} (\bibinfo{year}{1998}).

\bibitem[{\citenamefont{Rub{'i} and Mazur}(2000)}]{mazur00}
\bibinfo{author}{\bibfnamefont{J.}~\bibnamefont{Rub{'i}}} \bibnamefont{and}
  \bibinfo{author}{\bibfnamefont{P.}~\bibnamefont{Mazur}},
  \bibinfo{journal}{Physica A} \textbf{\bibinfo{volume}{276}},
  \bibinfo{pages}{477} (\bibinfo{year}{2000}).

\bibitem[{\citenamefont{Bedeaux and Mazur}(2001)}]{mazur01}
\bibinfo{author}{\bibfnamefont{D.}~\bibnamefont{Bedeaux}} \bibnamefont{and}
  \bibinfo{author}{\bibfnamefont{P.}~\bibnamefont{Mazur}},
  \bibinfo{journal}{Physica A} \textbf{\bibinfo{volume}{298}},
  \bibinfo{pages}{81} (\bibinfo{year}{2001}).

\bibitem[{\citenamefont{Rub{'i} and P{'e}rez-Madrid}(2001)}]{rubi01}
\bibinfo{author}{\bibfnamefont{J.}~\bibnamefont{Rub{'i}}} \bibnamefont{and}
  \bibinfo{author}{\bibfnamefont{A.}~\bibnamefont{P{'e}rez-Madrid}},
  \bibinfo{journal}{Physica A} \textbf{\bibinfo{volume}{298}},
  \bibinfo{pages}{177} (\bibinfo{year}{2001}).

\bibitem[{\citenamefont{Mackey}(1989)}]{mcm89rmp}
\bibinfo{author}{\bibfnamefont{M.~C.} \bibnamefont{Mackey}},
  \bibinfo{journal}{Rev. Mod. Phys.} \textbf{\bibinfo{volume}{61}},
  \bibinfo{pages}{981} (\bibinfo{year}{1989}).

\end{thebibliography}
\end{document}